\documentclass[preprint,aps,prd,a4paper,tightenlines,nofootinbib,showpacs]{revtex4-1}
\usepackage{amsmath,amssymb,graphicx,mathrsfs,color}
\usepackage{bm} 
\usepackage{feynmp}
\usepackage{ifpdf}
\ifpdf
\fi
\makeatletter
\def\endfmffile{%
  \fmfcmd{\p@rcent\space the end.^^J%
          end.^^J%
          endinput;}%
  \if@fmfio
    \immediate\closeout\@outfmf
  \fi
  \IfFileExists{\thefmffile.mp}{\immediate\write18{mpost \thefmffile}}{}
  \let\thefmffile\relax
}
\makeatother
\newcommand{\ds}{\displaystyle}

\newcommand{\nn}{\nonumber\\}

\newcommand{\la}{\langle}
\newcommand{\ra}{\rangle}

\newcommand{\bGamma}{\bar{\Gamma}}

\newcommand{\ben}{\begin{displaymath}}
\newcommand{\een}{\end{displaymath}}
\newcommand{\be}{\begin{equation}}
\newcommand{\ee}{\end{equation}}
\newcommand{\bea}{\begin{eqnarray}}
\newcommand{\eea}{\end{eqnarray}}

\newcommand{\A}{\alpha}

\newcommand{\G}{\Gamma}
\newcommand{\bG}{\bar{\Gamma}}

\newcommand{\q}{\bar{q}}

\newcommand{\bc}{\begin{center}}
\newcommand{\ec}{\end{center}}

\newcommand{\eqn}[1]{\label{#1}}
\newcommand{\eq}[1]{Eq.~(\ref{#1})}
\newcommand{\eqs}[1]{Eqs.~(\ref{#1})}
\newcommand{\fign}[1]{\label{#1}}
\newcommand{\fig}[1]{Fig.~\ref{#1}}

\begin{document}
\title{Covariant equations for the tetraquark and more}
\author{A. N. Kvinikhidze}
\affiliation{A.\ Razmadze Mathematical 
Institute, Georgian Academy of Sciences, M.\ Aleksidze St.\ 1, 380093 Tbilisi, Georgia}
\email{sasha\_kvinikhidze@hotmail.com}
\author{B. Blankleider}
\affiliation{
School of Chemical and Physical Sciences,
 Flinders University, Bedford Park, SA 5042, Australia}
\email{boris.blankleider@flinders.edu.au}

\date{\today}

\begin{abstract}
We derive covariant equations for a system of two quarks and two antiquarks where the effect of quark-antiquark annihilation is taken into account. 
In our approach, only pair-wise interactions are retained, while all possibilities of overcounting are excluded by (i) keeping terms in the kernel that are consistent with a meson-meson and diquark-antidiquark substructure, and (ii) introducing 4-body equations with a novel structure that specifically avoids the generation of overcounted terms. The resulting tetraquark bound state equations are given for the case of general two-body interactions, and for the specific case of separable interactions that lead to a description of the tetraquark in terms of meson-meson and diquark-antidiquark degrees of freedom where the effects of quark-antiquark annihilation is included. The inclusion of $2q2\q$- and $q\q$-channel coupling in our approach enables a wide variety of applications of our equations to other processes within the $2q2\q$ system, and to other 2-particle plus 2-antiparticle systems.

\end{abstract}

\maketitle

\section{Introduction}

  The present paper is motivated by recent studies by Heupel, Eichmann, Popovici and Fischer (HEPF) \cite{heupel,popovici} of  tetraquark bound states in the framework of covariant four-body equations based on continuum quantum field theory (QFT). These authors described the underlying two-quark two-antiquark ($2q2\q$) dynamics  in terms of  meson, diquark, and antidiquark degrees of freedom, and for this purpose used the four-body equations of Khvedelidze and Kvinikhidze (KK) \cite{kvin4q}. The equations of KK are exact  in the pair-interaction approximation, but  are valid, strictly speaking, only for systems like $4q$ where annihilation does not take place; as such, they can describe the $2q2\q$ system only if the effect of $q\q$ pair annihilation is neglected. The purpose of the present paper is to derive equations for the system of two quarks and two antiquarks where $q\q$ pair annihilation is taken into account.

In the context of continuum QFT, the derivation of covariant equations for few-body systems is an important but nontrivial task, as one routinely encounters the notorious problem of overcounting of Feynman diagrams. Such overcounting problems have been solved in the last two decades for a number of cases \cite{kvin4q,piNN,overcount,gaug1,gaug2,gaug-spect}. One of these is the case of four-quark equations where even the seemingly natural task of summing pair interaction kernels leads to overcounting \cite{kvin4q,heupel}. Likewise, overcounting provides a major challenge when formulating $2q2\q$ equations where $q\q$ annihilation is  taken into account. In this paper we have solved the overcounting problem by (i) keeping terms in the kernel that are consistent with the meson-meson ($MM$) and diquark-antidiquark ($D\bar{D}$)  coupled channels approach of HEPF, and (ii) introducing new coupled channel 4-body equations whose form is explicitly constructed to avoid overcounting.
In this way we have derived $2q2\q$ equations whose kernels encode the process of $q\bar q$ annihilation. 

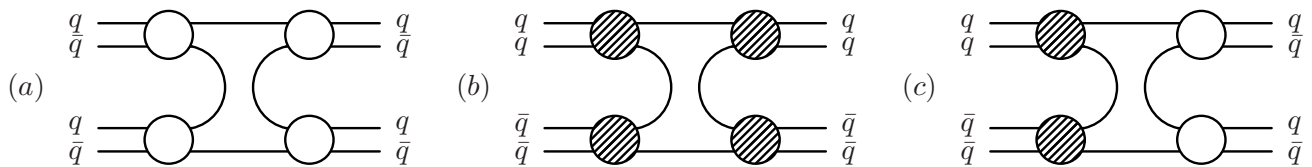
\begin{figure}[b]
\begin{center}

\begin{fmffile}{miss}
\[
\hspace{-5mm}
(a)\hspace{7mm}
\parbox{37mm}{
\begin{fmfgraph*}(37,17)\fmfkeep{MM}
\fmfstraight
\fmfleftn{f}{12}\fmfrightn{i}{12}
\fmf{plain}{i12,f12}
\fmf{plain}{f1,i1}
\fmf{phantom}{f10,t1,t2,t22,t3,t4,tm,t6,t7,t88,t8,t9,i10}
\fmf{phantom}{f3,b1,b2,b22,b3,b4,bm,b6,b7,b88,b8,b9,i3}
\fmf{phantom}{f11,tt1,tt2,tt22,tt3,tt4,ttm,tt6,tt7,tt88,tt8,tt9,i11}
\fmf{phantom}{f2,bb1,bb2,bb22,bb3,bb4,bbm,bb6,bb7,bb88,bb8,bb9,i2}
\fmffreeze
\fmf{plain}{t3,f10}
\fmf{plain}{i10,t7}
\fmf{plain}{f3,b3}
\fmf{plain}{b7,i3}
\fmf{plain,left=.8,tension=.5}{b7,t7}
\fmf{plain,right=.8,tension=.5}{b3,t3}
\fmfv{decor.shape=circle,decor.filled=empty, decor.size=18}{tt22}
\fmfv{decor.shape=circle,decor.filled=empty, decor.size=18}{bb22}
\fmfv{decor.shape=circle,decor.filled=empty, decor.size=18}{tt88}
\fmfv{decor.shape=circle,decor.filled=empty, decor.size=18}{bb88}
\fmfv{label=$q$,l.a=0}{i12}
\fmfv{label=$\q$,l.a=0}{i10}
\fmfv{label=$\q$,l.a=-10}{i1}
\fmfv{label=$q$,l.a=10}{i3}
\fmfv{label=$q$,l.a=180}{f12}
\fmfv{label=$\q$,l.a=180}{f10}
\fmfv{label=$\q$,l.a=190}{f1}
\fmfv{label=$q$,l.a=170}{f3}
\end{fmfgraph*}}
\hspace{1cm}
(b)\hspace{7mm}
\parbox{37mm}{
\begin{fmfgraph*}(37,17)\fmfkeep{DD}
\fmfstraight
\fmfleftn{f}{12}\fmfrightn{i}{12}
\fmf{plain}{i12,f12}
\fmf{plain}{f1,i1}
\fmf{phantom}{f10,t1,t2,t22,t3,t4,tm,t6,t7,t88,t8,t9,i10}
\fmf{phantom}{f3,b1,b2,b22,b3,b4,bm,b6,b7,b88,b8,b9,i3}
\fmf{phantom}{f11,tt1,tt2,tt22,tt3,tt4,ttm,tt6,tt7,tt88,tt8,tt9,i11}
\fmf{phantom}{f2,bb1,bb2,bb22,bb3,bb4,bbm,bb6,bb7,bb88,bb8,bb9,i2}
\fmffreeze
\fmf{plain}{t3,f10}
\fmf{plain}{i10,t7}
\fmf{plain}{f3,b3}
\fmf{plain}{b7,i3}
\fmf{plain,left=.8,tension=.5}{b7,t7}
\fmf{plain,right=.8,tension=.5}{b3,t3}
\fmfv{decor.shape=circle,decor.filled=shaded, decor.size=18}{tt22}
\fmfv{decor.shape=circle,decor.filled=shaded, decor.size=18}{bb22}
\fmfv{decor.shape=circle,decor.filled=shaded, decor.size=18}{tt88}
\fmfv{decor.shape=circle,decor.filled=shaded, decor.size=18}{bb88}
\fmfv{label=$q$,l.a=0}{i12}
\fmfv{label=$q$,l.a=0}{i10}
\fmfv{label=$\q$,l.a=-10}{i1}
\fmfv{label=$\q$,l.a=10}{i3}
\fmfv{label=$q$,l.a=180}{f12}
\fmfv{label=$q$,l.a=180}{f10}
\fmfv{label=$\q$,l.a=190}{f1}
\fmfv{label=$\q$,l.a=170}{f3}
\end{fmfgraph*}} \hspace{1cm}
(c)\hspace{7mm}
\parbox{37mm}{
\begin{fmfgraph*}(37,17)\fmfkeep{DM}
\fmfstraight
\fmfleftn{f}{12}\fmfrightn{i}{12}
\fmf{plain}{i12,f12}
\fmf{plain}{f1,i1}
\fmf{phantom}{f10,t1,t2,t22,t3,t4,tm,t6,t7,t88,t8,t9,i10}
\fmf{phantom}{f3,b1,b2,b22,b3,b4,bm,b6,b7,b88,b8,b9,i3}
\fmf{phantom}{f11,tt1,tt2,tt22,tt3,tt4,ttm,tt6,tt7,tt88,tt8,tt9,i11}
\fmf{phantom}{f2,bb1,bb2,bb22,bb3,bb4,bbm,bb6,bb7,bb88,bb8,bb9,i2}
\fmffreeze
\fmf{plain}{t3,f10}
\fmf{plain}{i10,t7}
\fmf{plain}{f3,b3}
\fmf{plain}{b7,i3}
\fmf{plain,left=.8,tension=.5}{b7,t7}
\fmf{plain,right=.8,tension=.5}{b3,t3}
\fmfv{decor.shape=circle,decor.filled=shaded, decor.size=18}{tt22}
\fmfv{decor.shape=circle,decor.filled=shaded, decor.size=18}{bb22}
\fmfv{decor.shape=circle,decor.filled=empty, decor.size=18}{tt88}
\fmfv{decor.shape=circle,decor.filled=empty, decor.size=18}{bb88}
\fmfv{label=$q$,l.a=0}{i12}
\fmfv{label=$\q$,l.a=0}{i10}
\fmfv{label=$\q$,l.a=-10}{i1}
\fmfv{label=$q$,l.a=10}{i3}
\fmfv{label=$q$,l.a=180}{f12}
\fmfv{label=$q$,l.a=180}{f10}
\fmfv{label=$\q$,l.a=190}{f1}
\fmfv{label=$\q$,l.a=170}{f3}
\end{fmfgraph*}}
\]
\end{fmffile}   
\vspace{-1mm}

\caption{\fign{miss}  Examples of terms involving $q\q$ annihilation  contributing to the tetraquark amplitude within a model involving only meson, diquark and antidiquark constituents: (a) $MM$ scattering, (b) $D\bar{D}$ scattering,  (c) $D\bar{D}\leftarrow MM$ transition. Such terms are not taken into account when the covariant four-body equations of KK \cite{kvin4q} are used to describe the $2q2\q$ system.}
\end{center}
\end{figure}

Although our equations,  \eqs{PsiAT}, do not depend on the form of the two-body interaction, the case of separable interactions corresponding to bound-state mesons, diquarks, and antidiquarks, is of special interest as it provides a description of the tetraquark in terms $MM$ and $D\bar{D}$ degrees of freedom, similar to that of HEPF, yet where $q\q$ annihilation is taken into account through inclusion of  processes like those illustrated in \fig{miss}. Our final equations, \eqs{eq-Phi-MD1-sym}, not only allow one to assess the contribution of non-exotic $q\q$ states to the makeup of tetraquarks, but they can more generally serve as a tool for identifying the tetraquark, meson molecule, or hybrid states; and given the covariant field-theoretical setting, they can offer better insights into the underlying dynamics of the strong interaction.

Apart from tetraquarks,  our equations can describe the scattering processes of usual mesons ($q\bar q$ bound states). They are also suitable for the construction of scattering amplitudes corresponding to all possible processes in the system of two nucleons and two anti-nucleons. The results of the paper are also useful for detailed studies of the non-exotic $q\bar q$ bound states  \cite{Fischer:2014xha}; namely, the two-body $q\bar q$ Bethe-Salpeter (BS)  kernel can be constructed as a solution of the $2q2\bar{q}$ equations, corresponding to an infinite sum of physically meaningful Feynman diagrams in the $q\bar q$ kernel.  Analogous studies could also be considered in the system of two electrons and two positrons.

\section{ Derivation}

In this section we present a derivation of four-body equations for  the $2q2\q$ system where quark annihilation is taken into account. For clarity of presentation, we treat the quarks as distinguishable, as all the necessary antisymmetrization can be performed at the end  of the derivation (Green functions and t matrices need to be summed with respect to permutations of either initial or final state quark/antiquark quantum numbers using anti-symmetrising factors of $-1$).
\bigskip

The four-body Green function $G$ and the corresponding t matrix $X$, defined by
\be
G=G_0+G_0XG_0,  \eqn{GX}
\ee
 satisfy Dyson equations which relate them to the four-body interaction kernel $K$:
\begin{subequations} \eqn{GK}
\begin{align}
G&=G_0+G_0KG,\\[2mm]
X&=K+KG_0X ,
\end{align}
\end{subequations}
where $G_0$ is the free four-body Green function. Following the notation of Ref.\ \cite{heupel}, we assign labels 1,2
to the quarks and 3,4 to the antiquarks. The kernel $K$ can be formally expressed as
\be
K=K_2+K_3
\ee
where $K_2$ consists of only pair-wise interactions, and $K_3$ consists of all other contributions, necessarily involving three- and four-body forces.
One can then write $K_2$ as a sum of three terms whose structure is illustrated in \fig{K}, and correspondingly expressed as
\be
K_2=\sum_{aa'} K_{aa'}  \eqn{pair}
\ee
where the index $a\in \left\{12,13,14,23,24,34\right\}$, enumerates six possible pairs of particles, 
and the double index, $aa'\in \left\{(12,34), (13,24),(14,23)\right\}$,  enumerates three possible two pairs of particles.
Thus $ K_{aa'}$ describes the part of
the four-body kernel where all interactions are switched off except those within the pairs $a$ and $a'$.
\begin{figure}[b]
\begin{center}
\begin{fmffile}{K}
\[
K_2\hspace{2mm}=\hspace{7mm}
\parbox{20mm}{
\begin{fmfgraph*}(20,15)
\fmfstraight
\fmfleft{f4,f3,f2,f1}\fmfright{i4,i3,i2,i1}
\fmf{plain,tension=1.3}{i1,v1,f1}
\fmf{plain,tension=1.3}{i2,v2,f2}
\fmf{plain,tension=1.3}{i3,v3,f3}
\fmf{plain,tension=1.3}{i4,v4,f4}
\fmffreeze
\fmf{phantom,tension=1.3}{v1,v12,v2}
\fmfv{d.s=circle,d.f=empty,d.si=14}{v12}
\fmf{phantom,tension=1.3}{v3,v34,v4}
\fmfv{d.s=circle,d.f=empty,d.si=14}{v34}
\fmfv{label=$1$,l.a=0}{i1}
\fmfv{label=$2$,l.a=0}{i2}
\fmfv{label=$3$,l.a=0}{i3}
\fmfv{label=$4$,l.a=0}{i4}
\fmfv{label=$q$,l.a=180}{f1}
\fmfv{label=$q$,l.a=180}{f2}
\fmfv{label=$\q$,l.a=180}{f3}
\fmfv{label=$\q$,l.a=180}{f4}
\end{fmfgraph*}}
\hspace{7mm}
+
\hspace{7mm}
\parbox{20mm}{
\begin{fmfgraph*}(20,15)
\fmfstraight
\fmfleft{f4,f2,f3,f1}\fmfright{i4,i2,i3,i1}
\fmf{plain,tension=1.3}{i1,v1,f1}
\fmf{plain,tension=1.3}{i2,v2,f2}
\fmf{plain,tension=1.3}{i3,v3,f3}
\fmf{plain,tension=1.3}{i4,v4,f4}
\fmffreeze
\fmf{phantom,tension=1.3}{v1,v13,v3}
\fmfv{d.s=circle,d.f=empty,d.si=14}{v13}
\fmf{phantom,tension=1.3}{v2,v24,v4}
\fmfv{d.s=circle,d.f=empty,d.si=14}{v24}
\fmfv{label=$1$,l.a=0}{i1}
\fmfv{label=$2$,l.a=0}{i2}
\fmfv{label=$3$,l.a=0}{i3}
\fmfv{label=$4$,l.a=0}{i4}
\fmfv{label=$q$,l.a=180}{f1}
\fmfv{label=$q$,l.a=180}{f2}
\fmfv{label=$\q$,l.a=180}{f3}
\fmfv{label=$\q$,l.a=180}{f4}
\end{fmfgraph*}}
\hspace{7mm}
+
\hspace{7mm}
\parbox{20mm}{
\begin{fmfgraph*}(20,15)
\fmfstraight
\fmfleft{f3,f2,f4,f1}\fmfright{i3,i2,i4,i1}
\fmf{plain,tension=1.3}{i1,v1,f1}
\fmf{plain,tension=1.3}{i2,v2,f2}
\fmf{plain,tension=1.3}{i3,v3,f3}
\fmf{plain,tension=1.3}{i4,v4,f4}
\fmffreeze
\fmf{phantom,tension=1.3}{v1,v14,v4}
\fmfv{d.s=circle,d.f=empty,d.si=14}{v14}
\fmf{phantom,tension=1.3}{v2,v23,v3}
\fmfv{d.s=circle,d.f=empty,d.si=14}{v23}
\fmfv{label=$1$,l.a=0}{i1}
\fmfv{label=$2$,l.a=0}{i2}
\fmfv{label=$3$,l.a=0}{i3}
\fmfv{label=$4$,l.a=0}{i4}
\fmfv{label=$q$,l.a=180}{f1}
\fmfv{label=$q$,l.a=180}{f2}
\fmfv{label=$\q$,l.a=180}{f3}
\fmfv{label=$\q$,l.a=180}{f4}
\end{fmfgraph*}}
\]
\end{fmffile}   
\vspace{3mm}

\caption{\fign{K}  Structure of the four-body kernel $K_2$ where only two-body forces are included. Each of the summed diagrams corresponds to the term $K_{aa'}$ of \eq{pair}, where index $a$ ($a'$) is given by the numerical labels of the two top (bottom) quark ($q$) or antiquark ($\q$)  lines. Note that the precise mathematical meaning of each diagram is given by \eq{Kaa}. }
\end{center}
\end{figure}

In order to express $K_{aa'}$ in terms of the two-body kernels $K_a$, we make use of the Green function $G_{aa'}$ that corresponds to the sum of all Feynman diagrams where the pair $a$ is disconnected from the pair $a'$:
\be
G_{aa'}=G_aG_{a'}  \eqn{product}
\ee
where $G_a$ is the two-body Green function for particle pair $a$.
Both $G_a$ and the corresponding two-body t matrix $X_a$, defined by
\be
G_a=G_a^0+G_a^0X_aG_a^0,   \eqn{Ga-Xa}
\ee
 satisfy Dyson equations which relate them to the two-body interaction kernel $K_a$:
\begin{subequations} \eqn{GaKa}
\begin{align}
G_a&=G_a^0+G_a^0K_aG_a, \eqn{Ga-Ka}\\[2mm]
X_a&=K_a+K_aG_a^0X_a,   \eqn{Xa-Ka}
\end{align}
\end{subequations}
where $G_a^0$ is the free Green function for pair $a$. Similarly, both $G_{aa'}$ and the corresponding four-body t matrix $X_{aa'}$, defined by 
\be
G_{aa'}=G_0+G_0 X_{aa'}G_0,   \eqn{Gaa-Xaa}
\ee
satisfy Dyson equations which relate them to the four-body interaction kernel $K_{aa'}$:
\begin{subequations} \eqn{GaaKaa}
\begin{align}
G_{aa'}&=G_0+G_0K_{aa'}G_{aa'}, \eqn{Gaa-Kaa}\\[2mm]
X_{aa'}&=K_{aa'}+K_{aa'}G_0X_{aa'}.   \eqn{Xaa-Kaa}
\end{align}
\end{subequations}
Then using \eq{product}, \eq{Ga-Ka}, and \eq{Gaa-Kaa}
one obtains\footnote{In 4-body expressions we shall suppress factors of $G_a^0{}^{-1}$ associated with  non-interacting pairs in the $a$ channel. For example,   \eq{Kaa} is shorthand for $\ds K_{aa'}=K_a G_{a'}^0{}^{-1}+K_{a'}G_{a}^0{}^{-1}-K_aK_{a'} $ }
\be
K_{aa'}=K_a +K_{a'}-K_aK_{a'} .\eqn{Kaa}
\ee
In the case of $qq$ or $\q\q$ channels ($\A=12$,  $34)$, $K_a$ is the sum of two-body irreducible diagrams, all of which are connected. However, in the case of a $q\bar{q}$ channel $(\A=13, 14, 23, 24)$, $K_a$ also contains a disconnected part which corresponds to the annihilation (creation) of the $q\bar{q}$ pairs into (from) vacuum in the initial (final) states.
 This disconnected part  of $K_a$ can be derived from \eq{Ga-Ka} given that the same disconnectedness is present in $G_a$ in the form of the product of two single quark Green functions corresponding to the independent propagation of $q$ and $\q$ in the t-channel (this disconnected part of $G_a$ should not be confused with the free Green function $G^0_a$, which corresponds to the independent propagation of $q$ and $\q$ in the s-channel).  In other words, the $q\bar{q}$ t matrix $X_a$ has a disconnected part $A_a$ which  consists of two u-turned quark lines corresponding to the annihilation (creation) of the $q\bar{q}$ pairs in the initial (final) state, as illustrated in \fig{Td}.  Such disconnected parts are present in the $2q2\bar{q}$ system, and are not taken into account by the 4-body equations of Ref.\ \cite{kvin4q} (these equation were developed to describe 4-body systems like $4q$, where there are no annihilation channels). 

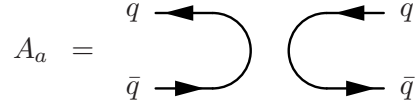
\begin{figure}[t]
\begin{center}
\begin{fmffile}{Ku}
\[
A_a\hspace{2mm}=\hspace{7mm}
\parbox{20mm}{
\begin{fmfgraph*}(30,10)
\fmfstraight
\fmfleft{f2,f1}\fmfright{i2,i1}
\fmftopn{t}{5}\fmfbottomn{b}{5}
\fmf{fermion}{t2,t1}
\fmf{fermion}{t5,t4}
\fmf{fermion}{b1,b2}
\fmf{fermion}{b4,b5}
\fmf{plain,left=1,tension=.5}{b4,t4}
\fmf{plain,right=1,tension=.5}{b2,t2}
\fmfv{label=$q$,l.a=180}{t1}
\fmfv{label=$\q$,l.a=180}{b1}
\fmfv{label=$q$,l.a=0}{t5}
\fmfv{label=$\q$,l.a=0}{b5}

\end{fmfgraph*}}
\]
\end{fmffile}   
\vspace{3mm}

\caption{\fign{Td}  Disconnected part of the $q\q$ t matrix $X_a$. Left arrows indicate particles, as labelled,  while right arrows indicate the corresponding antiparticles.}
\end{center}
\end{figure}

\subsection{Exact $2q2\q$ equations}\label{general}

Exact $2q2\q$ equations in the pairwise approximation can be obtained by analogy with the derivation for the covariant pion-two-nucleon ($\pi NN$) system \cite{Kvinikhidze:1993bn} where inclusion of pion absorption leads to a corresponding  overcounting  problem, as noted above. The procedure is to first expose  two-body $q\bar q$ cuts in the four-body Green function, as was done in Eq.\ (31) of Ref.\ \cite{Kvinikhidze:1993bn} where $N N$ cuts were exposed. The remainining $q\bar q$ irreducible part of the  $2q2\bar q$ Green function will then satisfy the $4q$ equations of Refs.\  \cite{kvin4q,heupel}. Details of this derivation will be presented elsewhere.

\subsection{Approximate $2q2\q$ equations}

Here we derive $2q2\bar q$ equations by modifying KK's $4q$ equations in such a way that disconnected 2-body kernels $A_a$ are included without any occurrence of overcounting.  Although this way may not be efficient for derivation of exact  $2q2\bar q$ equations, it suits well the nature of the approximations used in Ref.\ \cite{heupel} where only two-meson (MM) and diquark-antidiquark ($D\bar D$) states are exposed in the equations.
It is worth noting that the approach taken here is very different from the one used in Ref.\ \cite{piNN} to derive the $\pi NN$ equations.

We begin with Eqs.\ (\ref{GK}-\ref{Kaa}). The difference from the $4q$ case is that the $q\bar{q}$ kernels, $K_a$, contain disconnected parts which correspond to the annihilation of $q\bar{q}$ pairs into vacuum. Inclusion of these disconnected parts leads to an important difference between the $2q2\q$ formulation and the one for the $4q$ system: the "pair interaction approximation" where the full 4-body kernel $K$ is equated with the pairwise kernel $K_2$ of \eq{pair}, by itself, does not make sense in the $2q2\q$ case due to a double-counting problem in the corresponding Green function. In the exact $2q2\bar q$ equations described above, 3- and 4-body force  counterterms need to be included  in order to cancel the double-counted terms generated by iteration of the pair-interaction kernels - that is why discarding 3- and 4-body forces is not allowed in this setting. Here we show another way of avoiding this double-counting: we work out how to keep only that part of the pair interaction kernel which is physically meaningful on the one hand side, and that does not generate double-counted terms on the another.

We consider the two-body correlations t matrix $X_{aa'}$ defined by \eq{Gaa-Xaa}. Using \eq{product}, \eq{Ga-Xa}, and \eq{Gaa-Xaa}, one can express $X_{aa'}$ in terms of the two-body t matrices $X_a$ and  $X_{a'}$ as
\be
X_{aa'}=X_a+X_{a'}+X_aX_{a'} . \eqn{kern-Xaa'}
\ee
Writing
\be
X_a=T_a+A_a,
\ee
where $T_a$ and $A_a$ are the connected and disconnected parts of $X_a$, respectively, we note that  it is  $T_a$ which corresponds to the physical 2-body scattering amplitude, while the disconnected part $A_a$ contributes to the physical 4-body amplitude where it describes $q\bar{q}$ annihilation into vacuum. The explicit possibilities for $X_a$ are:
\begin{align}\label{Ta+Aa}
X_{12}&=T_{12}
\nn
X_{34}&=T_{34}
\nn
X_{13}&=T_{13}+A_{13}
\nn
X_{14}&=T_{14}+A_{14}
\nn
X_{23}&=T_{23}+A_{23}
\nn
X_{24}&=T_{24}+A_{24}.
\end{align}
Note that $A_a$ is non-zero only in $q\bar{q}$ subspace;  amplitude $A_{23}$, for example, is illustrated in \fig{A23} and explicitly given by
\be
A_{23}(k_2,k_3,p_2,p_3)=-\delta(p_2-p_3)S^{-1}(k_2)S^{-1}(p_2) \eqn{A23e}
\ee
where the momenta are assigned to the quark line direction, so that $p_2$  $(k_2)$ are the momenta of the incoming (outgoing) quarks, and 
$-p_3$ $(-k_3)$ are the momenta of the corresponding antiquarks. 
More specifically, one has the following quantum field theoretic definitions of the Green function quantities in the $23$ channel: 

\begin{figure}[t]
\begin{center}
\begin{fmffile}{A23}
\[
A_{23}(k_2,k_3,p_2,p_3) \hspace{2mm}=\hspace{7mm}
\parbox{20mm}{
\begin{fmfgraph*}(30,10)
\fmfstraight
\fmfleft{f2,f1}\fmfright{i2,i1}
\fmftopn{t}{5}\fmfbottomn{b}{5}
\fmf{fermion}{t2,t1}
\fmf{fermion}{t5,t4}
\fmf{fermion}{b1,b2}
\fmf{fermion}{b4,b5}
\fmf{plain,left=1,tension=.5}{b4,t4}
\fmf{plain,right=1,tension=.5}{b2,t2}
\fmfv{label=$k_2$,l.a=180}{t1}
\fmfv{label=$k_3$,l.a=180}{b1}
\fmfv{label=$p_2$,l.a=0}{t5}
\fmfv{label=$p_3$,l.a=0}{b5}

\end{fmfgraph*}}
\]
\end{fmffile}   
\vspace{3mm}

\caption{\fign{A23}  The amplitude $A_{23}$ (disconnected part of the $q\q$ t matrix $X_{23}$). With the initial (final) quark assigned momentum label $p_2$ ($k_2$), and corresponding antiquark assigned momentum $-p_3$ ($-k_3$), so that $p_2-p_3=0$, the expression for $A_{23}$ is given as in \eq{A23e}.}
\end{center}
\end{figure}
\begin{subequations}
\begin{align}
G_{23}(k_2,k_3,p_2,p_3)&=
\int e^{i(k_2y_2-k_3y_3-p_2x_2+p_3x_3)} \nn[-1mm]
&\hspace{1cm}\times\la\la 0|T q(y_2) \bar q(y_3)\bar q(x_2) q(x_3)|0\ra\ra
dy_2dy_3dx_2dx_3
\nn[2mm]
&=G_{23}^0+G_{23}^0(T_{23}+A_{23})G_{23}^0
\\[5mm]
G_{23}^0(k_2,k_3,p_2,p_3)
&=
\int e^{i(k_2y_2-p_2x_2)}\la\la 0|T q(y_2) \bar q(x_2)|0\ra\ra dy_2dx_2\nn
&\times
\int e^{i(-k_3y_3+p_3x_3)}
\la\la 0|Tq(x_3)\bar q(y_3)|0\ra\ra dy_3dx_3
\nn[2mm]
&=S(p_2)\delta(k_2-p_2)S(p_3)\delta(k_3-p_3)
\\[5mm]
\left[G^0_{23}A_{23}G^0_{23} \right](k_2,k_3,p_2,p_3)
&=
-\int e^{i(k_2y_2-k_3y_3)}\la\la 0|T q(y_2) \bar q(y_3)|0\ra\ra dy_2dx_2\nn
&\times
\int e^{i(-p_2x_2+p_3x_3)}
\la\la 0|Tq(x_3)\bar q(x_2)|0\ra\ra dy_3dx_3
\nn
&=-S(k_2)\delta(k_2-k_3)S(p_2)\delta(p_2-p_3).  \eqn{G0A23G0}
\end{align}
\end{subequations}
Note that the minus sign in the definition of $A_{23}$ is due to Wick theorem, as can be seen easily from the fact that $A_{23}$ can be obtained from $G^0_{23}$ by switching outgoing quark ends, thus entailing a sign change.

The covariant equations of KK were derived for four-body systems, like $4q$, where the pair interactions are described by connected 2-body t matrices $T_a$; that is, for the case where the disconnected parts $A_a$ are equal to zero for all 2-body channels $a$. Using the model where only 2-body correlations are included, $K=K_2$ (which obviously has no double-counting problems arising from disconnected two-body kernels), KK showed that the resultant 4-body t matrix, denoted by $T$, can be expressed as

\be
T=\sum_{aa'}{\cal T}_{aa'}  \eqn{T}
\ee
where 
\be
{\cal T}_{aa'}= T_{aa'} + T_{aa'}G_0 ({\cal T}_{bb'}+{\cal T}_{cc'}),\hspace{1cm} aa'\neq bb'\neq cc' \neq aa'. \eqn{Fad-Iak}
\ee
 In \eq{Fad-Iak}, the amplitude $T_{aa'}$ is the t matrix corresponding to Green function $G_{aa'}$; that is, $T_{aa'}\equiv X_{aa'}$ for the special case where all $A_a=0$, in which case \eq{kern-Xaa'} is written as
\be
T_{aa'}=T_a+T_{a'}+T_aT_{a'} . \eqn{kern-Taa'}
\ee
In order to derive covariant equations for the four-body system $2q2\q$ where $q\q$ annihilation is included, we shall start with the KK equations, \eq{T}, \eq{Fad-Iak}, and \eq{kern-Taa'}, and examine the consequence of including the disconnected parts $A_a$ by simply making the replacements $T_a\rightarrow X_a = T_a+A_a$; that is, we examine the consequences of writing the full amplitude $X$, defined in \eq{GX}, as
\be
X=\sum_{aa'}{\cal X}_{aa'}  \eqn{X}
\ee
where 
\be
{\cal X}_{aa'}= X_{aa'} + X_{aa'}G_0 ({\cal X}_{bb'}+{\cal X}_{cc'}),\hspace{1cm} aa'\neq bb'\neq cc' \neq aa',  \eqn{FadX}
\ee
with amplitude $X_{aa'}$ being given by \eq{kern-Xaa'}. For this purpose it is useful to introduce amplitudes $A_{aa'}$ defined by
\be
X_{aa'} = T_{aa'} + A_{aa'}
\ee
where $T_{aa'}$ is defined by \eq{kern-Taa'},  so that
\be
A_{aa'}=A_a+A_{a'}+T_aA_{a'}+A_aT_{a'}+A_aA_{a'} . \eqn{kern-Aaa'}
\ee
From the outset, we shall discard the product of disconnected terms $A_aA_{a'}$ (consisting of $A_{13}A_{24}$ and $A_{14}A_{23}$), as they do not contribute to the physically meaningful part of the 4-body t matrix. Nevertheless, it  is apparent (as demonstrated below) that \eq{FadX} is still problematic as it suffers from double-counting problems. Here we propose to handle the overcounting problem by making an approximation that is consistent with the one used in Ref.\cite{heupel}; namely, we shall neglect terms in  \eq{kern-Taa'} and \eq{kern-Aaa'} that are linear in $T_a$, so that\footnote{In the appendix we discuss the overcounting problem from a broader perspective where we show how more general equations $2q2\q$ can be derived for the case where the linear terms $T_a$ are retained.}
\begin{subequations}
\eqn{kernapprox}
\begin{align}
T_{aa'} &\rightarrow T_a T_{a'} \eqn{happ} \\
A_{aa'} &\rightarrow A_a + A_{a'}
\end{align}
\end{subequations}
In this approximation the two-body correlation t matrices $X_{aa'}$ are modelled as
\be
X_{aa'} \rightarrow T_aT_{a'}+A_a+A_{a'},
\ee
or specifically,
\begin{subequations}\label{appr-Taa'}
\begin{align}
X_{12,34}& \rightarrow T_{12}T_{34}\\[1mm]
X_{13,24} & \rightarrow
T_{13}T_{24}+A_{13}+A_{24}\\[1mm]
X_{14,23}& \rightarrow
T_{14}T_{23}+A_{14}+A_{23}
\end{align}
\end{subequations}
The approximation of \eq{happ} was used in Ref.\ \cite{heupel}, and is based on the physically motivated assumption of the tetraquark being mainly a bound state of two mesons or of diquark-antidiquark pairs. Thus only two $q\bar q$ pair interaction t matrices $( T_{13,24}$ and $T_{14,23})$ are modified with respect to the paper of Ref.\  \cite{heupel} through the addition of disconnected parts $A_a$. 
Analysis of the double-counting problem inherent in \eq{FadX} provides additional support for the approximations of \eq{appr-Taa'}. 
For example, consider the term $ T_aA_{a'}$ which appears in  \eq{kern-Aaa'}. Such a term is illustrated in \fig{oc1}(a)  for the  case of $T_{14}A_{23}$ which would arise as part of the inhomogeneous term $X_{14,23}$ of \eq{FadX}. Yet already in the second iteration of \eq{FadX} there would be the term $X_{14,23} G_0 X_{12,34} G_0 X_{14,23}$ giving rise to an amplitude $A_{23} T_{12}T_{34}  A_{23}$, illustrated in \fig{oc1}(b), which is already contained $T_{14}A_{23}$.
Thus, instead of introducing three- and four-body forces to compensate double-counted terms, as prescribed by the exact approach, the approximations of  \eq{appr-Taa'} provide an alternative description where these compensating forces are effectively taken into account without going beyond a pair interaction model.

\begin{figure}[t]
\begin{center}
\begin{fmffile}{overcount}
\[
(a)\hspace{10mm}
\parbox{25mm}{
\begin{fmfgraph*}(25,17)
\fmfstraight
\fmfleftn{l}{7}\fmfrightn{r}{7}
\fmftopn{t}{5}\fmfbottomn{b}{5}
\fmf{phantom}{l2,c2,c3,c4,r2}
\fmf{phantom}{l3,d2,d3,d4,r3}
\fmf{phantom}{l5,e2,e3,e4,r5}
\fmf{phantom}{l6,f2,f3,f4,r6}
\fmffreeze
\fmf{plain}{t1,t5}
\fmf{plain}{l5,r5}
\fmf{plain}{b1,b2}\fmf{plain}{b4,b5}
\fmf{plain}{l3,d2}\fmf{plain}{d4,r3}
\fmf{plain,left=1,tension=.5}{d2,b2}
\fmf{plain,right=1,tension=.5}{d4,b4}

\fmfv{decor.shape=circle,decor.filled=full, decor.size=16}{f3}
\fmfv{label=$1$,l.a=0}{r7}
\fmfv{label=$4$,l.a=0}{r5}
\fmfv{label=$2$,l.a=0}{r3}
\fmfv{label=$3$,l.a=0}{r1}
\fmfv{label=$1$,l.a=180}{l7}
\fmfv{label=$4$,l.a=180}{l5}
\fmfv{label=$2$,l.a=180}{l3}
\fmfv{label=$3$,l.a=180}{l1}
\end{fmfgraph*}}
 \hspace{2cm}
(b)\hspace{10mm}
%
\parbox{45mm}{
\begin{fmfgraph*}(45,17)
\fmfstraight
\fmfleftn{l}{7}\fmfrightn{r}{7}
\fmftopn{t}{11}\fmfbottomn{b}{11}
\fmf{phantom}{l2,c2,c3,c4,c5,c6,c7,c8,c9,c10,r2}
\fmf{phantom}{l3,d2,d3,d4,d5,d6,d7,d8,d9,d10,r3}
\fmf{phantom}{l5,e2,e3,e4,e5,e6,e7,e8,e9,e10,r5}
\fmf{phantom}{l6,f2,f3,f4,f5,f6,f7,f8,f9,f10,r6}
\fmffreeze
\fmf{plain}{t1,t11}\fmf{plain}{b1,b11}

\fmf{plain}{l5,e2}\fmf{plain}{e10,r5}
\fmf{plain}{l3,d2}\fmf{plain}{d10,r3}
\fmf{plain,left=1,tension=.5}{e2,d2}
\fmf{plain,right=1,tension=.5}{e10,d10}

\fmf{plain}{e4,e8}\fmf{plain}{d8,d4}
\fmf{plain,right=1,tension=.5}{e4,d4}
\fmf{plain,left=1,tension=.5}{e8,d8}

\fmfv{decor.shape=circle,decor.filled=full,decor.size=16}{c6}
\fmfv{decor.shape=circle,decor.filled=full,decor.size=16}{f6}
\fmfv{label=$1$,l.a=0}{r7}
\fmfv{label=$2$,l.a=0}{r5}
\fmfv{label=$3$,l.a=0}{r3}
\fmfv{label=$4$,l.a=0}{r1}
\fmfv{label=$1$,l.a=180}{l7}
\fmfv{label=$2$,l.a=180}{l5}
\fmfv{label=$3$,l.a=180}{l3}
\fmfv{label=$4$,l.a=180}{l1}
\end{fmfgraph*}}
\]
\end{fmffile}   
\vspace{3mm}

\caption{\fign{oc1}  Example of overcounting resulting from a replacement $T_a \rightarrow X_a=T_a+A_a$ in the KK equations: (a) amplitude  $T_{14} A_{23}$ contained in the inhomogeneous  term $X_{14,23}$ of \eq{FadX},
(b)
amplitude $A_{23}  T_{12}T_{34}  A_{23}$ contained in the second iteration term $X_{14,23}G_0X_{12,34}G_0 X_{13,23}$. The amplitude in (b) is already contained in the amplitude in (a).}
\end{center}
\end{figure}
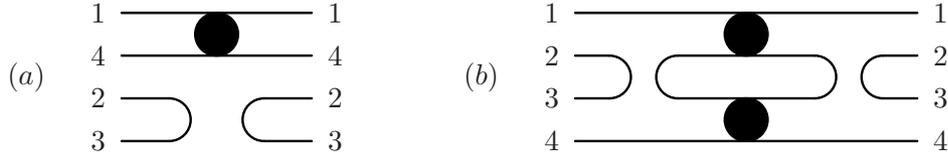

However, even with the approximations of \eq{appr-Taa'} implemented, \eq{FadX} still generates double-counted terms. Namely, the term $ A_{13}A_{14}$ generated in the first iteration, can be obtained from $ A_{13}$ by switching antiquark labels 3 and 4 in the initial state, as illustrated in \fig{oc2};  yet just this switching will be produced by antisymmetrisation of the solution of \eq{FadX}. Such troublesome double-counted terms can be avoided by modifying  \eq{FadX} in such a way that will not allow the kernels $A_{13}$ and $A_{14}$ to meet through the process of iteration.
\begin{figure}[b]
\begin{center}
\begin{fmffile}{AAovercount}
\[
\parbox{60mm}{
\begin{fmfgraph*}(60,17)
\fmfstraight
\fmfleftn{l}{4}\fmfrightn{r}{4}
\fmftopn{d}{10}\fmfbottomn{a}{10}
\fmf{phantom}{l3,c2,c3,c4,c5,c6,c7,c8,c9,r3}
\fmf{phantom}{l2,b2,b3,b4,b5,b6,b7,b8,b9,r2}
\fmffreeze
\fmf{fermion}{r1,a9}\fmf{plain}{a9,a6}\fmf{fermion}{a6,a4}\fmf{plain}{a4,a2}\fmf{fermion}{a2,l1}
\fmf{fermion}{l2,b2}\fmf{plain}{b2,b4}\fmf{fermion}{b4,b6}\fmf{fermion}{b9,r2}
\fmf{fermion}{l3,c2}\fmf{fermion}{c4,c6}\fmf{plain}{c6,c9}\fmf{fermion}{c9,r3}
\fmf{fermion}{r4,d9}\fmf{fermion}{d6,d4}\fmf{fermion}{d2,l4}
\fmf{plain,left=1,tension=.5}{d2,c2}
\fmf{plain,right=1,tension=.5}{d4,c4}
\fmf{plain,rubout=4,left=1,tension=.5}{d6,b6}
\fmf{plain,rubout=4,right=1,tension=.5}{d9,b9}
\fmfv{label=$1$,l.a=0}{r4}
\fmfv{label=$3$,l.a=0}{r3}
\fmfv{label=$4$,l.a=0}{r2}
\fmfv{label=$2$,l.a=0}{r1}
\fmfv{label=$1$,l.a=180}{l4}
\fmfv{label=$3$,l.a=180}{l3}
\fmfv{label=$4$,l.a=180}{l2}
\fmfv{label=$2$,l.a=180}{l1}
\end{fmfgraph*}}
\]
\end{fmffile}   
\end{center}
\vspace{-3mm}
\caption{\fign{oc2}  Example of overcounting inherent in \eq{FadX}. Illustrated is amplitude $A_{13}  A_{14}$, which is generated in the first iteration of  \eq{FadX}, but that is also obtained from $ A_{13}$ when the initial state antiquark labels 3 and 4 are interchanged through antisymmetrisation.}
\end{figure}
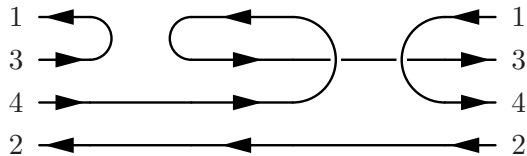
To this end we split the component amplitudes ${\cal X}_{aa'}$  into two parts
 \be
  {\cal X}_{aa'} =  {\cal T}_{aa'} +{\cal A}_{aa'}   \eqn{X=T+A}
\ee
and introduce the following coupled equations that replace those of \eq{FadX}:
\begin{subequations}\label{neweq}
\begin{align}
{\cal A}_{aa'}&= A_{aa'} + A_{aa'}G_0({\cal T}_{bb'}+{\cal T}_{cc'}),\\[2mm]
{\cal T}_{aa'}&= T_{aa'} + T_{aa'}G_0({\cal X}_{bb'}+{\cal X}_{cc'}),
\end{align}
\end{subequations}
where $aa'\neq bb'\neq cc' \neq aa'$. \eqs{neweq} provide the sought-after description of the covariant $2q2\q$ system where $q\q$ annihilation is taken into account in a way that is free from overcounting, and that is consistent with the two-meson and diquark-antidiquark model of Ref.\ \cite{heupel}.

The corresponding tetraquark bound state equations are
\be
\Psi= \sum_{aa'} \Psi_{aa'}  \eqn{Psi-sum}
\ee
where 
\be
\Psi_{aa'}= \Psi^T_{aa'} + \Psi^A_{aa'}
\ee
and
\begin{subequations}\label{PsiAT}
\begin{align}
\Psi^A_{aa'}&= G_0 A_{aa'}(\Psi^T_{bb'}+\Psi^T_{cc'}), \eqn{PsiA}\\[2mm]
\Psi^T_{aa'}&=  G_0 T_{aa'}(\Psi_{bb'}+\Psi_{cc'}), \eqn{PsiT}
\end{align}
\end{subequations}
where $aa'\neq bb'\neq cc' \neq aa'$.  To save on notation, we shall suppress writing factors of $G_0$. Then from \eqs{PsiAT} one obtains the following closed form equation for $\Psi^T_{aa'}$: 
\begin{align}
\Psi^T_{aa'}
&= T_{aa'}\left[(1+A_{cc'})\Psi^T_{bb'}+(1+A_{bb'})\Psi^T_{cc'}+(A_{bb'}+A_{cc'})\Psi^T_{aa'}\right]. \eqn{eq-Psi2}
\end{align}
One should note that the kernels $T_{aa'}$ are not compact as they contain singular $\delta$-functions corresponding to the pair $a(a')$ total 4-momentum conservation.
Similarly, the kernels $T_{aa'}A_{cc'}$ are not compact either as they involve  $\delta$-functions restricting the total momentum of some $q\bar q$ pairs to zero.
One should therefore iterate \eq{eq-Psi2} once to cast it in the form where the kernels are compact. The procedure of compactification is simpler if one uses the separable approximation for two-body t matrices:
\be
T_{a} = -\G_a D_a \bG_a  \eqn{Ta-sep}
\ee
where $D_a$ is the propagator for the bound particle in channel $a$  (diquark, antidiquark, or meson), and $\Gamma$ ($\bGamma$) is the vertex function for the particle's disintegration into (formation from) its quark  or antiquark constituents.\footnote{Note that our definitions of $\Gamma$ and $\bGamma$ are the ones often used for separable potentials, but differ from the ones used in Ref.\ \cite{heupel}. }  Showing explicit dependence on momentum variables, \eq{Ta-sep} can be expressed as
\be
T_{a}(p'_1 p'_2,p_1p_2)=-\G_a(p'_1p'_2)D_a(P)\bG_a(p_1p_2), \eqn{R-L1} 
\ee
where $P=p_1+p_2$ is the total off-mass-shell momentum of the bound particle. Substiting into \eq{eq-Psi2}  leads to the factorization of the $2q2\q$ bound state wave function as:
\begin{subequations}\label{Psi-simp1}
 \begin{align}
 \Psi^T_{aa'} &=  \G_a D_a  \G_{a'} D_{a'} \Phi_{aa'} \eqn{PsiTa}\\[1mm]
 \Phi_{aa'} &=  \bG_a \bG_{a'} \left[(1+A_{cc'})\Psi^T_{bb'}+(1+A_{bb'})\Psi^T_{cc'}+(A_{bb'}+A_{cc'})\Psi^T_{aa'}\right].  \eqn{PsiTb}
\end{align}
\end{subequations}
where $\Phi_{aa'}$ are the components of the $2q2\bar q$ bound state vertex function in $MM$ and $D\bar D$ space, i.e., $\Phi_{13,24}$ ($\Phi_{12,34}$) is the $MM\theta$  ($D\bar{D}\theta$) covariant vertex. As functions of momenta, \eq{PsiTa} can be written as
\be
\Psi^T_{aa'}(p,q,q',P)=\G_a(q,Q)D_a(Q)\G_{a'}(q',Q')D_{a'}(Q')\Phi_{aa'}(p,P)  \eqn{fact-1}
\ee
where $P$ is the $2q2\q$ bound state total momentum, $p$ is the
relative momentum between its respective constituents,
 $q, q'$ are
the relative momenta of the (anti-)diquarks and mesons,
$Q, Q'$ are their off-mass-shell momenta. 
Using \eq{PsiTa} in \eq{PsiTb} one obtains a closed set of equations for the bound state vertex functions:
\begin{align}
\Phi_{aa'}&=\bG_a \bG_{a'}  (1+A_{cc'}) \G_b \G_{b'} \ D_b D_{b'} \Phi_{bb'}\nn[1mm]
&+\bG_a \bG_{a'} (1+A_{bb'}) \G_c \G_{c'} \ D_c D_{c'}\Phi_{cc'}\nn[1mm]
&+\bG_a \bG_{a'} (A_{bb'}+A_{cc'}) \G_a  \G_{a'} \ D_a D_{a'} \Phi_{aa'},  \eqn{eq-Phi} 
\end{align}
where $aa'\neq bb'\neq cc' \neq aa'$. Although formally a set of three coupled equations, consideration of antisymmetry reduces \eqs{eq-Phi} to a set of two equations for the two components, $MM$ and $D\bar{D}$, of the tetraquark. To show this, we define the component vertex functions as
\begin{subequations}\label{D-M-comp}
\begin{align}
\Phi_D&=\Phi_{12,34} \\
\Phi_M&=\Phi_{13,24}=-\Phi_{14,23}.
\end{align}
\end{subequations}
The relation, $\Phi_{13,24}=-\Phi_{14,23}$, for the $MM$ component of the tetraquark follows from the antisymmetry of the diquark and antidiquark wave functions with respect to permutation of the quarks' quantum numbers:  $\Gamma_{12}=-\Gamma_{21}$ and $ \Gamma_{34}=-\Gamma_{43}$.  This antisymmetry property relates 
$MM\leftarrow D\bar D$ transition kernels to each other,
\be
 \bar\Gamma_{13} \bar\Gamma_{24}(1+A_{14,23})\Gamma_{12} \Gamma_{34}=-
 \bar\Gamma_{14} \bar\Gamma_{23}(1+A_{13,24})\Gamma_{12} \Gamma_{34}   \eqn{antisym}
\ee
which in turn can be used in  \eq{eq-Phi} to show $\Phi_{13,24}=-\Phi_{14,23}$.
  
The two equations for $\Phi_M$ and $\Phi_D$, mentioned above,  consist of two lines of \eqs{eq-Phi}, one corresponding to 
$aa'=13,24$, $bb'=12,34$, and $cc'=14,23$ (for which $A_{bb'}=0$), and another corresponding to $aa'=12,34$,  $bb'=13,24$, and $cc'=14,23$: 
\begin{subequations}\label{eq-Phi-MD1}
\begin{align}
\Phi_M&=\left( \bar\Gamma_{13} \bar\Gamma_{24}A_{14,23}\Gamma_{13} \Gamma_{24}- \bar\Gamma_{13} \bar\Gamma_{24}\Gamma_{14} \Gamma_{23}\right)MM\Phi_M\nn
&\hspace{2.2cm} +
 \bar\Gamma_{13} \bar\Gamma_{24}(1+A_{14,23})\Gamma_{12} \Gamma_{34} D \bar D\Phi_D, \\[2mm]
\Phi_D&=2 \bar\Gamma_{12} \bar\Gamma_{34}(1+A_{14,23})\Gamma_{13} \Gamma_{24}MM\Phi_M\nn
&\hspace{2.2cm} +4 \bar\Gamma_{12} \bar\Gamma_{34}A_{23}\Gamma_{12} \Gamma_{34}D\bar D\Phi_D
\end{align}
\end{subequations}
where we have used  $A_{14,23}=A_{14}+A_{23}$, and the following relations analogous to  \eq{antisym}:
\begin{subequations}
\begin{align}
 &\bG_{12} \bG_{34} (1+A_{14,23})\G_{13} \G_{24}-
 \bG_{12} \bG_{34}(1+A_{13,24})\G_{14} \G_{23}\nn
&\hspace{2cm}  =2 \bar\Gamma_{12} \bar\Gamma_{34}(1+A_{14,23})\Gamma_{13} \Gamma_{24}\\[2mm]
 &\bar\Gamma_{12} \bar\Gamma_{34} (A_{13,24}+A_{14,23}) \Gamma_{12} \Gamma_{34}\nn
 & \hspace{2cm} =
2 \bar\Gamma_{12} \bar\Gamma_{34}A_{13,24}\Gamma_{12} \Gamma_{34} = 4 \bar\Gamma_{12} \bar\Gamma_{34}A_{23}\Gamma_{12} \Gamma_{34}.
\end{align}
\end{subequations}
With respect to meson quantum numbers, \eqs{eq-Phi-MD1} admit both symmetric and antisymmetric solutions
because the kernels of \eqs{eq-Phi-MD1} do not change when the meson quantum numbers are swapped in initial and  final states simultaneously. To exclude the antisymmetric solutions, \eqs{eq-Phi-MD1} should be symmetrised with respect to meson quantum numbers.
Such symmetrization and the replacements (renormalization)
\be
 \Gamma_{12}\rightarrow \frac{1}{\sqrt{2}} \Gamma_{12},\hspace{2mm}
 \bar\Gamma_{12}\rightarrow \frac{1}{\sqrt{2}} \bar\Gamma_{12},\hspace{2mm}
\Gamma_{34}\rightarrow\frac{1}{\sqrt{2}}\Gamma_{34},\hspace{2mm}
\bar\Gamma_{34}\rightarrow\frac{1}{\sqrt{2}}\bar\Gamma_{34},\hspace{2mm}\Phi_D \rightarrow 2\Phi_D        \eqn{renormaliz}
\ee
cast the \eqs{eq-Phi-MD1} into a form where the symmetry with respect to the indistinguishable meson legs is manifest:
\begin{subequations}
\label{eq-Phi-MD1-sym}
\begin{align}
\Phi_M&=\sum_P\left( \bar\Gamma_{13} \bar\Gamma_{24}A_{14,23}\Gamma_{13} \Gamma_{24}- \bar\Gamma_{13} \bar\Gamma_{24}\Gamma_{14} \Gamma_{23}\right)\frac{MM}{2}\Phi_M \nn
&\hspace{3cm} +
 \bar\Gamma_{13} \bar\Gamma_{24} (1+A_{14,23}) \Gamma_{12} \Gamma_{34} D \bar D
\Phi_D, \\[3mm]
\Phi_D&= \bar\Gamma_{12} \bar\Gamma_{34} (1+A_{14,23}) \Gamma_{13} \Gamma_{24}\frac{MM}{2}\Phi_M\nn
&\hspace{3cm} + \bar\Gamma_{12} \bar\Gamma_{34}A_{23}\Gamma_{12} \Gamma_{34}D\bar D\Phi_D
\end{align}
\end{subequations}
where $\sum_P$ stands for the sum over meson legs' permutation in either initial or final state. Note the combinatorial normalization factor $1/2$ at each intermediate state of two indistinguishable mesons. To understand the renormalization in \eq{renormaliz}, we note that the symmetric \eq{eq-Phi-MD1-sym} could be obtained in the above derivation if one would renormalize the ansatz \eq{R-L1} for $qq$ and $\bar q\bar q$ amplitudes by factor $1/2$; for example, in the $qq$ case
\be
T_{12}(p_1'p'_2,p_1p_2)=-\frac{1}{2}\Gamma_{12}(p_1'p'_2)D(P)\bar\Gamma_{12}(p_1p_2).
\eqn{sym-ans} 
\ee
The $\Gamma_{12}$ extracted from \eq{sym-ans}, owing to the factor $1/2$,  is the correctly normalized vertex function of a diquark composed of indistinguishable quarks, in that $2T_{12}$ contains all diagrams of scattering of indistinguishable quarks; for example, in the one gluon exchange  approximation, $2T_{12}$ corresponds to the symmetric sum of two single gluon exchange diagrams.
 If, instead, the  ansatz of \eq{R-L1} and the corresponding \eqs{eq-Phi-MD1}
are slightly more convenient (as mentioned above), it is only because, for example, in the one gluon exchange  approximation kernel,  the vertex $\Gamma_{qq}$ is related to $\Gamma_{q\bar q}$ only by the substitution of an antiquark with a quark leg without a factor.
The reason is that the quark-quark and quark-antiquark scattering amplitudes, $T_{12}$ and $T_{13}$, satisfy the same equation, $T=K+KT$, where $K$ corresponds to a single diagram of one gluon exchange.

\subsection{Double counting problem}

Although the physically transparent form of our final equations for the vertex functions $\Phi_M$ and $\Phi_D$, \eqs{eq-Phi-MD1-sym},  should dispel any concerns that some important parts may still be missing or some parts still overcounted, there exists a rigorous way to check this.
To formulate exact QFT equations for few-body systems like $\pi NN$ and $2q2\q$ where some particles can be absorbed by others (e.g. $\pi$ by $N$), or pairs of particles can undergo annihilation (e.g. $q\bar q$), one starts with the general structure of the full few-body Green function that, in the case of the  $2q2\bar q$ system, is  manifested by the relation
\be
G^{(4)}=G^{(4)}_{ir}+G^{(4-2)}_{ir}G_0^{(2)-1}G^{(2)}G_0^{(2)-1}G^{(2-4)}_{ir},  \eqn{exact}
\ee
where $G^{(2)}$ is the full two-body $q\bar q$ Green function, and $G^{(4)}_{ir}$ is the $q\bar q$ irreducible part of the full  $2q2\bar q$ Green function $G^{(4)}$; further, $G^{(2-4)}_{ir}$ ($G^{(4-2)}_{ir}$) is  the sum of all $q\bar q$  irreducible diagrams of the Green function corresponding to the transition $q\bar q\leftarrow 2q2\bar q$ ($2q2\bar q\leftarrow q\bar q$). The main task is then  to express  $G^{(2-4)}_{ir}$ and $G^{(4-2)}_{ir}$  in terms of $G^{(4)}_{ir}$. To be consistent with the problem setting (which is to derive equations coupling the $q\q$ and $2q2\bar q$ channels), $G^{(2)}$ also should be expressed  in terms of $G^{(4)}_{ir}$ (thereby exposing the $2q2\bar q$ intermediate states in $G^{(2)}$).  For $G^{(2-4)}_{ir}$, this would normally be accomplished by isolating the last possible $2q2\q$ cut in $G^{(2-4)}_{ir}$, thereby splitting this amplitude into two parts:   $G^{(4)}_{ir}$ to the right of the cut, and a  $q\bar q\leftarrow 2q2\bar q$ amplitude that is both $q\q$ and $2q2\q$ irreducible, to the left of the cut. The problem is that such a "last  $2q2\q$ cut" is not unique, and special procedures need to be implemented to avoid consequent overcounting.  In just this way, QFT few-body equations were derived for the $\pi NN$ problem in Ref.\ \cite{piNN}.

We now show that our final equations, \eqs{eq-Phi-MD1-sym}, can be cast into the form specified by \eq{exact}. To begin, we rewrite \eqs{eq-Phi-MD1-sym} in the form
\begin{subequations}
\label{eq-Phi-sym-rewr}
\begin{align}
\Phi_M&= (\bar\Gamma_{13} \bar\Gamma_{24})^S\left(A_{23}-\frac{P_{34}}{2}\right)(\Gamma_{13} \Gamma_{24})^S\frac{MM}{2}\Phi_M+
(\bar\Gamma_{13} \bar\Gamma_{24})^S\left(\frac{1}{2}+A_{23}\right)
\Gamma_{12} \Gamma_{34} D \bar D\Phi_D, \\
\Phi_D&= \bar\Gamma_{12} \bar\Gamma_{34}\left(\frac{1}{2}+A_{23}\right)(\Gamma_{13} \Gamma_{24})^S\frac{MM}{2}\Phi_M
+ \bar\Gamma_{12} \bar\Gamma_{34}A_{23}\Gamma_{12} \Gamma_{34}D\bar D\Phi_D
\end{align}
\end{subequations}
where $(\bar\Gamma_{13} \bar\Gamma_{24})^S$ denotes the wave function of two indistinguishable mesons, so that 
\be\label{Ga-sym-index1}
(\Gamma_{13} \Gamma_{24})^S=\Gamma_{13}^{p} \Gamma_{24}^{k}+\Gamma_{13}^{k} \Gamma_{24}^{p},
\ee
where $p$ and $k$ are the meson momenta, and 
$P_{34}$ stands for permutation of antiquark legs 3 and 4, so that
\be
P_{34}\Gamma_{13}^p \Gamma_{24}^k=\Gamma_{14}^p \Gamma_{23}^k.
\ee
The set of \eqs{eq-Phi-sym-rewr} can be written in matrix form as
\be
\Phi=VG_0^M\Phi,  \eqn{homo-MM-1}
\ee
where 
\be
\Phi(p,k)=\left( \begin{array}{c} \Phi_M(p,k)\\[1mm]
 \Phi_D(p,k) \end{array} \right), \hspace{1cm}
G_0^M=\left( \begin{array}{cc}\frac{1}{2}MM  & 0 \\[1mm]
 0 &D\bar{D} \end{array} \right),  \eqn{PSI-S}
\ee
and $V$ is a $2\times 2$ matrix kernel corresponding to all four transitions between $MM$ and $D\bar D$ states. One can express $V$ as a sum $V=V_{q\bar q}+V_{2q2\bar q}$ where
  $V_{q\bar q}$ and $V_{2q2\bar q}$ are the parts of the kernel corresponding to $q\bar q$ and $2q2\bar q$ s-channel exchanges, respectively:
\begin{subequations}
\eqn{kernels}
\be
V_{q\bar q} =
\left( \begin{array}{cc} (\bar\Gamma_{13} \bar\Gamma_{24})^SA_{23}(\Gamma_{13} \Gamma_{24})^S &
(\bar\Gamma_{13} \bar\Gamma_{24})^SA_{23}\Gamma_{12} \Gamma_{34}
\\[2mm]
\bar\Gamma_{12} \bar\Gamma_{34}A_{23}(\Gamma_{13} \Gamma_{24})^S  & \bar\Gamma_{12} \bar\Gamma_{34}A_{23}\Gamma_{12} \Gamma_{34}
\end{array} \right),
\eqn{symm}
\ee
\be
V_{2q2\bar q} =\frac{1}{2}
\left( \begin{array}{cc} -(\bar\Gamma_{13} \bar\Gamma_{24})^SG_0P_{34}(\Gamma_{13} \Gamma_{24})^S &
(\bar\Gamma_{13} \bar\Gamma_{24})^SG_0\Gamma_{12} \Gamma_{34} 
\\[2mm]
\bar\Gamma_{12} \bar\Gamma_{34}G_0(\Gamma_{13} \Gamma_{24})^S  & 0
\end{array} \right).
\eqn{symm-4q}
\ee
\end{subequations}
Note that the propagator for four non-interacting quarks, $G_0$, is shown explicitly in \eq{symm-4q} whereas in \eqs{eq-Phi-sym-rewr} it is omitted for notational convenience. These kernels are illustrated diagrammatically in \fig{kernelsd}.
\begin{figure}
\begin{fmffile}{Vqq}
\[
(a)\hspace{1cm} V_{q\bar q} = \begin{pmatrix} \hspace{1mm}
\parbox{50mm}{
\begin{fmfgraph*}(37,17)\fmfkeep{MM}
\fmfstraight
\fmfleftn{f}{12}\fmfrightn{i}{12}
\fmf{plain}{i12,f12}
\fmf{plain}{f1,i1}
\fmf{phantom}{f10,t1,t2,t22,t3,t4,tm,t6,t7,t88,t8,t9,i10}
\fmf{phantom}{f3,b1,b2,b22,b3,b4,bm,b6,b7,b88,b8,b9,i3}
\fmf{phantom}{f11,tt1,tt2,tt22,tt3,tt4,ttm,tt6,tt7,tt88,tt8,tt9,i11}
\fmf{phantom}{f2,bb1,bb2,bb22,bb3,bb4,bbm,bb6,bb7,bb88,bb8,bb9,i2}
\fmffreeze
\fmf{plain}{t3,f10}
\fmf{plain}{i10,t7}
\fmf{plain}{f3,b3}
\fmf{plain}{b7,i3}
\fmf{plain,left=.8,tension=.5}{b7,t7}
\fmf{plain,right=.8,tension=.5}{b3,t3}
\fmfv{decor.shape=circle,decor.filled=empty,decor.size=18}{tt22}
\fmfv{decor.shape=circle,decor.filled=empty,decor.size=18}{bb22}
\fmfv{decor.shape=circle,decor.filled=empty,decor.size=18}{tt88}
\fmfv{decor.shape=circle,decor.filled=empty,decor.size=18}{bb88}
\fmfv{label=$q$,l.a=0}{i12}
\fmfv{label=$\q$,l.a=0}{i10}
\fmfv{label=$\q$,l.a=-10}{i1}
\fmfv{label=$q$,l.a=10}{i3}
\fmfv{label=$q$,l.a=180}{f12}
\fmfv{label=$\q$,l.a=180}{f10}
\fmfv{label=$\q$,l.a=190}{f1}
\fmfv{label=$q$,l.a=170}{f3}
\end{fmfgraph*}}
&
\parbox{47mm}{
\begin{fmfgraph*}(37,17)\fmfkeep{MD}
\fmfstraight
\fmfleftn{f}{12}\fmfrightn{i}{12}
\fmf{plain}{i12,f12}
\fmf{plain}{f1,i1}
\fmf{phantom}{f10,t1,t2,t22,t3,t4,tm,t6,t7,t88,t8,t9,i10}
\fmf{phantom}{f3,b1,b2,b22,b3,b4,bm,b6,b7,b88,b8,b9,i3}
\fmf{phantom}{f11,tt1,tt2,tt22,tt3,tt4,ttm,tt6,tt7,tt88,tt8,tt9,i11}
\fmf{phantom}{f2,bb1,bb2,bb22,bb3,bb4,bbm,bb6,bb7,bb88,bb8,bb9,i2}
\fmffreeze
\fmf{plain}{t3,f10}
\fmf{plain}{i10,t7}
\fmf{plain}{f3,b3}
\fmf{plain}{b7,i3}
\fmf{plain,left=.8,tension=.5}{b7,t7}
\fmf{plain,right=.8,tension=.5}{b3,t3}
\fmfv{decor.shape=circle,decor.filled=empty, decor.size=18}{tt22}
\fmfv{decor.shape=circle,decor.filled=empty, decor.size=18}{bb22}
\fmfv{decor.shape=circle,decor.filled=shaded, decor.size=18}{tt88}
\fmfv{decor.shape=circle,decor.filled=shaded, decor.size=18}{bb88}
\fmfv{label=$q$,l.a=0}{i12}
\fmfv{label=$q$,l.a=0}{i10}
\fmfv{label=$\q$,l.a=-10}{i1}
\fmfv{label=$\q$,l.a=10}{i3}
\fmfv{label=$q$,l.a=180}{f12}
\fmfv{label=$\q$,l.a=180}{f10}
\fmfv{label=$\q$,l.a=190}{f1}
\fmfv{label=$q$,l.a=170}{f3}
\end{fmfgraph*}}\hspace{2mm}\\[17mm]
 \hspace{1mm} \parbox{50mm}{
\begin{fmfgraph*}(37,17)\fmfkeep{DM}
\fmfstraight
\fmfleftn{f}{12}\fmfrightn{i}{12}
\fmf{plain}{i12,f12}
\fmf{plain}{f1,i1}
\fmf{phantom}{f10,t1,t2,t22,t3,t4,tm,t6,t7,t88,t8,t9,i10}
\fmf{phantom}{f3,b1,b2,b22,b3,b4,bm,b6,b7,b88,b8,b9,i3}
\fmf{phantom}{f11,tt1,tt2,tt22,tt3,tt4,ttm,tt6,tt7,tt88,tt8,tt9,i11}
\fmf{phantom}{f2,bb1,bb2,bb22,bb3,bb4,bbm,bb6,bb7,bb88,bb8,bb9,i2}
\fmffreeze
\fmf{plain}{t3,f10}
\fmf{plain}{i10,t7}
\fmf{plain}{f3,b3}
\fmf{plain}{b7,i3}
\fmf{plain,left=.8,tension=.5}{b7,t7}
\fmf{plain,right=.8,tension=.5}{b3,t3}
\fmfv{decor.shape=circle,decor.filled=shaded, decor.size=18}{tt22}
\fmfv{decor.shape=circle,decor.filled=shaded, decor.size=18}{bb22}
\fmfv{decor.shape=circle,decor.filled=empty, decor.size=18}{tt88}
\fmfv{decor.shape=circle,decor.filled=empty, decor.size=18}{bb88}
\fmfv{label=$q$,l.a=0}{i12}
\fmfv{label=$\q$,l.a=0}{i10}
\fmfv{label=$\q$,l.a=-10}{i1}
\fmfv{label=$q$,l.a=10}{i3}
\fmfv{label=$q$,l.a=180}{f12}
\fmfv{label=$q$,l.a=180}{f10}
\fmfv{label=$\q$,l.a=190}{f1}
\fmfv{label=$\q$,l.a=170}{f3}
\end{fmfgraph*}}
&
\parbox{47mm}{
\begin{fmfgraph*}(37,17)\fmfkeep{DD}
\fmfstraight
\fmfleftn{f}{12}\fmfrightn{i}{12}
\fmf{plain}{i12,f12}
\fmf{plain}{f1,i1}
\fmf{phantom}{f10,t1,t2,t22,t3,t4,tm,t6,t7,t88,t8,t9,i10}
\fmf{phantom}{f3,b1,b2,b22,b3,b4,bm,b6,b7,b88,b8,b9,i3}
\fmf{phantom}{f11,tt1,tt2,tt22,tt3,tt4,ttm,tt6,tt7,tt88,tt8,tt9,i11}
\fmf{phantom}{f2,bb1,bb2,bb22,bb3,bb4,bbm,bb6,bb7,bb88,bb8,bb9,i2}
\fmffreeze
\fmf{plain}{t3,f10}
\fmf{plain}{i10,t7}
\fmf{plain}{f3,b3}
\fmf{plain}{b7,i3}
\fmf{plain,left=.8,tension=.5}{b7,t7}
\fmf{plain,right=.8,tension=.5}{b3,t3}
\fmfv{decor.shape=circle,decor.filled=shaded, decor.size=18}{tt22}
\fmfv{decor.shape=circle,decor.filled=shaded, decor.size=18}{bb22}
\fmfv{decor.shape=circle,decor.filled=shaded, decor.size=18}{tt88}
\fmfv{decor.shape=circle,decor.filled=shaded, decor.size=18}{bb88}
\fmfv{label=$q$,l.a=0}{i12}
\fmfv{label=$q$,l.a=0}{i10}
\fmfv{label=$\q$,l.a=-10}{i1}
\fmfv{label=$\q$,l.a=10}{i3}
\fmfv{label=$q$,l.a=180}{f12}
\fmfv{label=$q$,l.a=180}{f10}
\fmfv{label=$\q$,l.a=190}{f1}
\fmfv{label=$\q$,l.a=170}{f3}
\end{fmfgraph*}} \hspace{2mm}
\end{pmatrix}
\]
\end{fmffile}   
\vspace{3mm}

\begin{fmffile}{Vqqqq}
\[
(b)\hspace{1cm} V_{2q2\q} = \frac{1}{2} \begin{pmatrix} \hspace{1mm}
\parbox{50mm}{
\begin{fmfgraph*}(37,17)\fmfkeep{MM}
\fmfstraight
\fmfleftn{f}{12}\fmfrightn{i}{12}
\fmf{plain}{i12,f12}
\fmf{plain}{f1,i1}
\fmf{phantom}{f10,t1,t2,t22,t3,t4,tm,t6,t7,t88,t8,t9,i10}
\fmf{phantom}{f3,b1,b2,b22,b3,b4,bm,b6,b7,b88,b8,b9,i3}
\fmf{phantom}{f11,tt1,tt2,tt22,tt3,tt4,ttm,tt6,tt7,tt88,tt8,tt9,i11}
\fmf{phantom}{f2,bb1,bb2,bb22,bb3,bb4,bbm,bb6,bb7,bb88,bb8,bb9,i2}
\fmffreeze
\fmf{plain}{t3,f10}
\fmf{plain}{i10,t7}
\fmf{plain}{f3,b3}
\fmf{plain}{b7,i3}
\fmf{plain,tension=.5}{b7,t3}
\fmf{plain,tension=.5}{b3,t7}
\fmfv{decor.shape=circle,decor.filled=empty, decor.size=18}{tt22}
\fmfv{decor.shape=circle,decor.filled=empty, decor.size=18}{bb22}
\fmfv{decor.shape=circle,decor.filled=empty, decor.size=18}{tt88}
\fmfv{decor.shape=circle,decor.filled=empty, decor.size=18}{bb88}
\fmfv{label=$q$,l.a=0}{i12}
\fmfv{label=$\q$,l.a=0}{i10}
\fmfv{label=$\q$,l.a=-10}{i1}
\fmfv{label=$q$,l.a=10}{i3}
\fmfv{label=$q$,l.a=180}{f12}
\fmfv{label=$\q$,l.a=180}{f10}
\fmfv{label=$\q$,l.a=190}{f1}
\fmfv{label=$q$,l.a=170}{f3}
\end{fmfgraph*}}
&
\parbox{47mm}{
\begin{fmfgraph*}(37,17)\fmfkeep{MD}
\fmfstraight
\fmfleftn{f}{12}\fmfrightn{i}{12}
\fmf{plain}{i12,f12}
\fmf{plain}{f1,i1}
\fmf{phantom}{f10,t1,t2,t22,t3,t4,tm,t6,t7,t88,t8,t9,i10}
\fmf{phantom}{f3,b1,b2,b22,b3,b4,bm,b6,b7,b88,b8,b9,i3}
\fmf{phantom}{f11,tt1,tt2,tt22,tt3,tt4,ttm,tt6,tt7,tt88,tt8,tt9,i11}
\fmf{phantom}{f2,bb1,bb2,bb22,bb3,bb4,bbm,bb6,bb7,bb88,bb8,bb9,i2}
\fmffreeze
\fmf{plain}{t3,f10}
\fmf{plain}{i10,t7}
\fmf{plain}{f3,b3}
\fmf{plain}{b7,i3}
\fmf{plain,tension=.5}{b7,t3}
\fmf{plain,tension=.5}{b3,t7}
\fmfv{decor.shape=circle,decor.filled=empty, decor.size=18}{tt22}
\fmfv{decor.shape=circle,decor.filled=empty, decor.size=18}{bb22}
\fmfv{decor.shape=circle,decor.filled=shaded, decor.size=18}{tt88}
\fmfv{decor.shape=circle,decor.filled=shaded, decor.size=18}{bb88}
\fmfv{label=$q$,l.a=0}{i12}
\fmfv{label=$q$,l.a=0}{i10}
\fmfv{label=$\q$,l.a=-10}{i1}
\fmfv{label=$\q$,l.a=10}{i3}
\fmfv{label=$q$,l.a=180}{f12}
\fmfv{label=$\q$,l.a=180}{f10}
\fmfv{label=$\q$,l.a=190}{f1}
\fmfv{label=$q$,l.a=170}{f3}
\end{fmfgraph*}} \hspace{2mm}\\[17mm]
 \hspace{1mm} \parbox{50mm}{
\begin{fmfgraph*}(37,17)\fmfkeep{DM}
\fmfstraight
\fmfleftn{f}{12}\fmfrightn{i}{12}
\fmf{plain}{i12,f12}
\fmf{plain}{f1,i1}
\fmf{phantom}{f10,t1,t2,t22,t3,t4,tm,t6,t7,t88,t8,t9,i10}
\fmf{phantom}{f3,b1,b2,b22,b3,b4,bm,b6,b7,b88,b8,b9,i3}
\fmf{phantom}{f11,tt1,tt2,tt22,tt3,tt4,ttm,tt6,tt7,tt88,tt8,tt9,i11}
\fmf{phantom}{f2,bb1,bb2,bb22,bb3,bb4,bbm,bb6,bb7,bb88,bb8,bb9,i2}
\fmffreeze
\fmf{plain}{t3,f10}
\fmf{plain}{i10,t7}
\fmf{plain}{f3,b3}
\fmf{plain}{b7,i3}
\fmf{plain,tension=.5}{b7,t3}
\fmf{plain,tension=.5}{b3,t7}
\fmfv{decor.shape=circle,decor.filled=shaded, decor.size=18}{tt22}
\fmfv{decor.shape=circle,decor.filled=shaded, decor.size=18}{bb22}
\fmfv{decor.shape=circle,decor.filled=empty, decor.size=18}{tt88}
\fmfv{decor.shape=circle,decor.filled=empty, decor.size=18}{bb88}
\fmfv{label=$q$,l.a=0}{i12}
\fmfv{label=$\q$,l.a=0}{i10}
\fmfv{label=$\q$,l.a=-10}{i1}
\fmfv{label=$q$,l.a=10}{i3}
\fmfv{label=$q$,l.a=180}{f12}
\fmfv{label=$q$,l.a=180}{f10}
\fmfv{label=$\q$,l.a=190}{f1}
\fmfv{label=$\q$,l.a=170}{f3}
\end{fmfgraph*}}
&
0\hspace{2mm}
\end{pmatrix}
\]
\end{fmffile}   
\vspace{3mm}

\caption{\fign{kernelsd} Diagrammatic representation of the kernels of \eqs{kernels}: (a) kernels with $q\q$ intermediate states, as given explicitly in \eq{symm-2q}, (b) kernels with $2q2\q$ intermediate states, as given explicitly by \eq{symm-4q}.}
\end{figure}
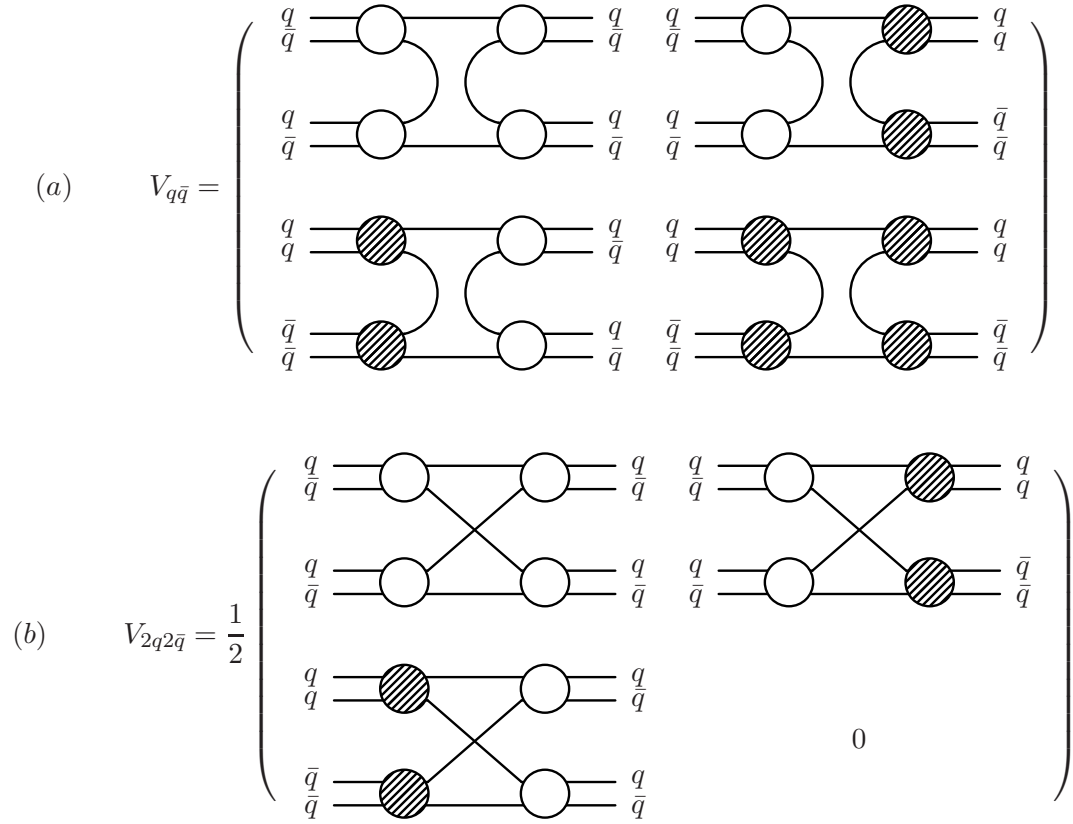

The inhomogeneous equation for the $MM$-$D\bar{D}$ Green function, $G$, corresponding to the homogeneous  \eq{homo-MM-1} is
\be
G=G_0^M+G_0^M(V_{q\bar q}+V_{2q2\bar q})G.  \eqn{simpl-eq-1}
\ee 
It can be written in the form
\be
G=G_{ir}+G_{ir}V_{q\bar q}G.  \eqn{7-1}
\ee
where $G_{ir}$ is the sum of all $q\bar q$ irreducible terms in Green function G, and itself satisfies the equation
\be
G_{ir}=G_0^M+G_0^M V_{2q2\bar q}G_{ir}. \eqn{8-1}
\ee
Using \eq{G0A23G0}, which we shall write in the current 4-body context as $A_{23}=-S_{23}G_{14}^0S_{23}$, \eq{symm} can be written as
\begin{align}
V_{q\bar q} &=-
\left( \begin{array}{cc} (\bar\Gamma_{13} \bar\Gamma_{24})^SS_{23}G_{14}^0S_{23}(\Gamma_{13} \Gamma_{24})^S &
(\bar\Gamma_{13} \bar\Gamma_{24})^SS_{23}G_{14}^0S_{23}\Gamma_{12} \Gamma_{34}
\\[2mm]
\bar\Gamma_{12} \bar\Gamma_{34}S_{23}G_{14}^0S_{23}(\Gamma_{13} \Gamma_{24})^S  & \bar\Gamma_{12} \bar\Gamma_{34}S_{23}G_{14}^0S_{23}\Gamma_{12} \Gamma_{34}
\end{array} \right)
\nn[2mm]
&=-
\left( \begin{array}{cc} N_{MM}G_{14}^0\bar N_{MM} &
N_{MM}G_{14}^0\bar N_{D\bar D}
\\[2mm]
N_{D\bar D}G_{14}^0\bar N_{MM}  &N_{D\bar D} G_{14}^0\bar N_{D\bar D}
\end{array} \right),
\eqn{symm-2q}
\end{align}
where the repeated indices 2 and 3, which stand for quantum numbers of the second quark and the third antiquark, are mute, i.e. they are summation indices. Therefore, for example, the expression
\be
N_{MM}=(\bar\Gamma_{13} \bar\Gamma_{24})^SS_{23}
\ee
is the amplitude of the transition of quark 1 and antiquark 4 to two mesons, and the propagator $S_{23}$ corresponds to the internal exchanged quark  line. Similarly 
\be
N_{D\bar D}=\bar\Gamma_{12} \bar\Gamma_{34}S_{23}
\ee
is the amplitude of the transition of quark 1 and antiquark 4 to a diquark-antidiquark pair.
The kernel of \eq{symm-2q}  thus consists of terms of the form $N_i (p',k')G^0_{14}\bar N_j(p,k)$ as illustrated in \fig{kernelsd}(a), where initial and final
$MM$ or $D\bar{D}$ states are separated only by a 2-body $q\bar q$ intermediate state.
 The matrix of \eq{symm-2q} can be written in a compact symbolic form as a direct product of column ($N$) and row ($\bar N$) matrices: 
\be
V_{q\bar q}=-NG_{q\q}^0\bar N
\eqn{VqqG-1}
\ee
where $G_{q\q}^0=G_{14}^0$ , and
\begin{subequations}
\begin{align}
N &= \begin{pmatrix} 
N_{MM}\\[2mm]
 N_{D\bar D}\end{pmatrix} 
 =\begin{pmatrix} (\bar\Gamma_{13} \bar\Gamma_{24})^SS_{23}\\[2mm]
\bar\Gamma_{12} \bar\Gamma_{34}S_{23}\end{pmatrix},\\[3mm]
\bar N &=
\begin{pmatrix} \bar N_{MM} & \bar N_{D\bar D}\end{pmatrix} =
\begin{pmatrix} S_{23} (\Gamma_{13} \Gamma_{24})^S &
S_{23}\Gamma_{12} \Gamma_{34}
\end{pmatrix}.
\end{align}
\end{subequations}
Then from \eq{7-1} and \eq{8-1} we get
\be
G=G_{ir}+G_{ir}NG_{q\bar q}\bar N G_{ir}  \eqn{9-1}
\ee
where $G_{q\bar q}$ is the $q\bar q$ Green function which contains all $q\bar q$ intermediate states, and is itself determined by equation
\be
G_{q\bar q}=G_{q\q}^0+G_{q\q}^0\left(\bar N G_{ir}N\right)G_{q\bar q} . \eqn{10-1}
\ee
Here $\bar N G_{ir}N$ is the $q\bar q$ interaction potential. 
In \eq{9-1} the $q\bar q$ cuts ($G^0_{q\q}$) are exposed via Green function $G_{q\bar q}$, as specified by \eq{10-1}.
With \eq{9-1}, we have obtained the  realisation of \eq{exact} for the particular case of the separable approximation of \eq{R-L1}. It is interesting to note that in this approximate case, we have been able to derive equations for the $2q2\q$ system where $q\q$ annihilation is included, but without having to face the above mentioned ambiguity of  the  last $2q2\bar q$ cut in $G^{(2-4)}$.   The point is that the term $\bar N G_{ir}$  appearing in \eq{9-1} contains within it just such a last $2q2\bar q$ cut that is not unique, yet $\bar N$ is determined unambiguously in \eq{9-1}. 
Careful analysis of this approximate model may thus lead us to the solution of this well-known problem of the general case; in particular, it may be possible to deduce a criterion which helps one to make an unambiguous choice of the very last cut, such that the double-counting problem is avoided. Finally, we note that 
the decomposition of \eq{9-1} allows one to see whether something is overcounted, or what may be missing, in the initial approximate description provided by \eqs{PsiAT}, and how it may be improved.

\section{discussion}

In this paper we have derived covariant equations for the $2q2\q$ system where $q\q$ annihilation is taken into account. This has been achieved 
in a model where the kernel consists only of terms that allow for a description in terms of $MM$ and $D\bar{D}$ degrees of freedom in the case where the separable approximations of \eq{R-L1} are used for the  two-body interactions. We find it encouraging that the parts of the kernel that are neglected in this model (namely, $T_a$, $T_a A_{a'}$, and $A_aA_{a'}$, see \eqs{kernapprox}) are just the ones that would cause overcounting if retained. The parts of the kernel retained in the model (namely $T_aT_{a'}$ and $A_a$), can still cause overcounting when two disconnected terms, $A_{13}$ and $A_{14}$, are allowed to meet through iteration. To stop this from happening, we have introduced  4-body equations with a novel structure designed specifically to prevent this type of overcounting.

For the general case of two-body interactions, our equations for the tetraquark bound state are given by \eqs{PsiAT}. For two-body separable interactions, as specified by \eq{R-L1}, our equations are expressed in terms of $MM$ and $D\bar D$ degrees of freedom, and presented first for distinguishable particles,  \eq{eq-Phi}. Taking into account the antisymmetry of identical quarks and antiquarks, but  without reference to the symmetry of the two meson states, the equations reduce  from 3 coupled equations down to two, as given by \eqs{eq-Phi-MD1}. Finally, with the mesons symmetrised,  we obtain \eqs{homo-MM-1}.

Our \eqs{homo-MM-1}  reduce to those of  Ref.\ \cite{heupel} if we eliminate the effect of $q\q$ annihilation by setting
$V_{q\bar q}=0$, or equivalently, by setting $A_{aa'}=0$ in \eqs{eq-Phi-MD1}. The kernels involving $A_{aa'}$ correspond to quark box diagrams, as in \fig{kernelsd}(a), where two-body $q\bar q$  intermediate states are incorporated. In this way the two-body $q\bar q$ component contributions are buried in the kernels of \eqs{eq-Phi-MD1}, even though they are written in terms of only meson and diquark degrees of freedom.
Adding these box diagrams does not complicate the tetraquark equations of Ref.~\cite{heupel} in the sense they are one-loop diagrams, just like the kernels in Ref.\ \cite{heupel}. The complication is that one gets two equations instead of one.
Also, including the box diagrams makes the model complete up to the one loop level since all other effects involve
two and more loops in the kernel.

It is worth pointing out that the $MM$-$D\bar D$ picture of a tetraquark follows from the separable approximation for the input two-body scattering amplitudes, and that the addition of the box diagrams is not something that is beyond this approximation:  one only adds some disconnected parts in $q\bar q$ channels to make the equations applicable to the $2q 2\bar q$ system. This addition restores missing topologies, and is not part of the dynamics: all the dynamics is encoded in the two-body scattering amplitudes $T_a$.

With our equations, it will be possible to ascertain the importance of $q\q$ annihilation in the description of the tetraquark, in a quantitative way.

\appendix*
\section{Overcounting in \eq{FadX}}

In this appendix we explain in more detail how the overcounting problem encountered in Eq. (20), is solved.
The simplest double-counted term appears already in the first iteration of \eq{FadX}, resulting in iterated  disconnected terms $A_{ij}$:
\be
 X_{13,24}(X_{14,23}+X_{12,34}) = A_{13}A_{14}+...
\ee
whose double-counting property was illustrated in \fig{oc2}.  Similarly, we illustrated in \fig{oc1} how the second iteration, $X_{14,23} X_{12,34}X_{14,23}$, generates the term  $A_{23} T_{12}T_{34}A_{23}$ which is already contained in the amplitude $ T_{14}A_{23}$ which forms part of the inhomogeneous term $X_{14,23}$. Terms of the type $ T_aA_{a'}$ were discarded in the approximate kernels of  
\eqs{appr-Taa'} just  to avoid  such double-counting.

However, it is noteworthy that the term $A_{23} T_{12}T_{34}A_{23}$, by itself, involves double-counting in the case of exact  $ T_{12}$ and $T_{34}$. This is evident from \fig{oc1}(b) where the interaction between quark 1 and antiquark 4 is of the form given in \fig{T14oc}. Overcounting will occur if the amplitudes $T_{12}$ and $T_{34}$ contain t-channel exchanges of interacting $q\q$ pairs. On the other hand, in the often used rainbow-ladder approximation for the two-body t matrices (which is a factorisation assumption for the t matrices in the s-channel) no terms are double-counted. Indeed
\be
A_{23} T_{12}T_{34}A_{23}=A_{23}(p'_2,p'_3,p_2,p_3)\int (dk)S(k'_2,k'_3)T_{12}(p_1'k'_2,p_1k_2)T_{34}(k'_3p'_4,k_3p_4)S(k_2,k_3)    \eqn{T12T34}
\ee
where $S(k_2,k_3)=\delta (k_2-k_3)S(k_2)$ is a quark propagator, the integration over four momenta, $(dk)=dk'_2tk'_3dk_2dk_3$, is reduced to a one loop 4-momentum integral upon the use of  4-momentum conservation $\delta$-functions, including one coming from the scattering amplitude, $T_{12}(p_1'k'_2,p_1k_2)=T_{12}(p_1'k'_2,p_1k_2)\delta(p_1'+k'_2-p_1-k_2)$. The rainbow-ladder approximation implies the factorization in the s-channel:
\be
T_{12}(p_1'p'_2,p_1p_2)=-\Gamma(p_1'p'_2)D(P)\bar\Gamma(p_1p_2),\hspace{1cm}P=p_1+p_2. \eqn{R-L}
\ee
Note that the same factorisation approximation for the two-body t matrices in the t-channel,
\be
T_{12}(p_1'p'_2,p_1p_2)=-\Gamma(p_1'p_1)M(p_1'-p_1)\bar\Gamma(p'_2p_2)
\ee
 would lead to a double-counting in  \eq{T12T34}. Indeed, this double-counting can be seen in the unphysical second order pole at
$(p_1'-p_1)^2=m_M^2$. 
\begin{figure}[t]
\begin{center}
\begin{fmffile}{T14oc}
\[
\parbox{30mm}{
\begin{fmfgraph*}(30,17)
\fmfstraight
\fmfleftn{l}{7}\fmfrightn{r}{7}
\fmftopn{t}{5}\fmfbottomn{b}{5}
\fmf{phantom}{l2,c2,c3,c4,r2}
\fmf{phantom}{l3,d2,d3,d4,r3}
\fmf{phantom}{l5,e2,e3,e4,r5}
\fmf{phantom}{l6,f2,f3,f4,r6}
\fmffreeze
\fmf{fermion}{t5,tc,t1}\fmf{fermion}{b1,bc,b5}

\fmf{fermion,right=1,tension=.5}{f3,c3}
\fmf{fermion,right=1,tension=.5}{c3,f3}

\fmfv{decor.shape=circle,decor.filled=full, decor.size=16}{c3}
\fmfv{decor.shape=circle,decor.filled=full, decor.size=16}{f3}
\fmfv{label=$1$,l.a=0}{r7}
\fmfv{label=$4$,l.a=0}{r1}
\fmfv{label=$1'$,l.a=180}{l7}
\fmfv{label=$4'$,l.a=180}{l1}

\end{fmfgraph*}}
\]
\end{fmffile}   
\vspace{3mm}

\caption{\fign{T14oc}  Form of the interaction between quark 1 and antiquark 4 inside the amplitude $A_{23}  T_{12}T_{34}  A_{23}$.}
\end{center}
\end{figure}
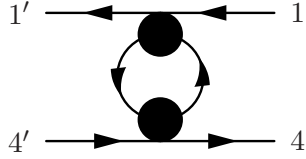

Because of the overcounting just discussed, we drop both the  $A_aA_{a'}$ and $T_aA_{a'}$ terms from the kernel $A_{aa'}$ as defined by   \eq{kern-Aaa'}; in this way,  we specify all the kernels as
\begin{subequations}
\label{kernels1}
\begin{align}
X_{aa'}&=T_{aa'}+A_{aa'} \\
T_{aa'}&=T_a+T_{a'}+T_aT_{a'}\\
A_{aa'} &=A_a+A_{a'}.  \eqn{A1}
\end{align}
\end{subequations}
Specifically, the kernels $X_{aa'}$ are given in the stated approximation as 
\begin{subequations}
\label{rich-appr-Taa'}
\begin{align}
X_{12,34}&=T_{12}+T_{34}+T_{12}T_{34}
\\
X_{13,24}
&=
T_{13}+T_{24}+T_{13}T_{24}+A_{13}+A_{24}
\\
X_{14,23}
&=
T_{14}+T_{23}+T_{14}T_{23}+A_{14}+A_{23}
\end{align}
\end{subequations}
These approximate kernels still generate double-counting once they are iterated via \eq{FadX}. In particular, iteration leads to
(i) the term $ A_{13}A_{14}$ which can be obtained from $ A_{13}$ by switching antiquark 3 and 4 legs in the initial state, but this term will be produced by antisymmetrisation of the solution of \eq{FadX}, (ii) the part $A_{23} T_{12}A_{23}$ of the second iteration,  $A_{23} X_{12,34}A_{23}$, which leads to an $A_{23}$ type term with an overdressed first quark line. Such overcounting can be avoided by modifying  \eq{FadX} in such a way that will prevent troublesome pairs of kernels, like $A_{13}$ and $A_{14}$, $A_{23}$ and $T_{12}$, etc., ever meeting each other when the equations are iterated.

To this end we split the kernels $T_{aa'}$ into two parts, specified as
\begin{subequations}
\begin{align}
T_{aa'}&=T^1_{aa'}+T^2_{aa'}\\[1mm]
T^1_{aa'} &=T_a+T_{a'}\\[1mm]
T^2_{aa'} &=T_aT_{a'},
\end{align}
\end{subequations}
and correspondingly, express the amplitude ${\cal X}_{aa'}$ of \eq{FadX} as
\begin{subequations}
\label{amps1}
\begin{align}
{\cal X}_{aa'}&={\cal T}_{aa'}+{\cal A}_{aa'} \\[1mm]
{\cal T}_{aa'}&={\cal T}^1_{aa'}+{\cal T}^2_{aa'}
\end{align}
\end{subequations}
With these definitions, the $2q2\q$ amplitude $X$ is given by
\be
X=\sum_{aa'} {\cal X}_{aa'},  \eqn{X=Xaa'} 
\ee
where the modified equations for the components ${\cal X}_{aa'}$  are given by
\begin{subequations}
\label{modified-eq}
\begin{align}
{\cal A}_{aa'}&= A_{aa'} + A_{aa'}\left({\cal T}^2_{bb'}+{\cal T}^2_{cc'}\right)
\\[1mm]
{\cal T}^1_{aa'}&= T^1_{aa'} + T^1_{aa'}\left({\cal T}_{bb'}+{\cal T}_{cc'}\right)
\\[1mm]
{\cal T}^2_{aa'}&= T^2_{aa'} + T^2_{aa'}\left({\cal X}_{bb'}+{\cal X}_{cc'}\right)
\end{align}
\end{subequations}
where $aa'\neq bb'\neq cc' \neq aa'$. 
\eqs{modified-eq} are obtained by expressing \eq{FadX} symbolically in terms of the above component amplitudes as
\begin{subequations}
\label{symb-eq-Fad-Iak}
\begin{align}
{\cal A}&= A + A \left(\underline{{\cal T}}^1+{\cal T}^2 +\underline{{\cal A}} \right),
\\
{\cal T}^1&= T^1 + T^1 \left( {\cal T}^1+{\cal T}^2 +\underline{{\cal A}} \right),
\\
{\cal T}^2&= T^2 + T^2 \left({\cal T}^1+{\cal T}^2 + {\cal A} \right),
\end{align} 
\end{subequations}
and discarding the underlined  terms. 
The term $A{\cal A}$ is discarded because it generates the problematic term $A_{13}A_{14}$ discussed above. The term $A{\cal T}^1$ is discarded because all the terms it generates in the second iteration of  \eqs{symb-eq-Fad-Iak},
\be
A{\cal T}^1\rightarrow AT^1({\cal T}^1+{\cal T}^2+{\cal A})
\rightarrow AT^1(T^1+ T^2+A)
\ee
 suffer doublecounting. For example, the part, $A_{23} T_{12}A_{23}$, of $ AT^1A$  is a $A_{23}$ type term with overdressed first quark line. The terms, $AT^1(T^1+ T^2)$, involve subdiagrams with a product of two two-body scattering amplitudes, $T^{(2)}G_0^{(2)}T^{(2)}$, similar to that of \fig{T14oc}. The term $ T^1{\cal A}\sim T_{14}A_{23}$ is discarded because the same term is partially obtained in $A{\cal T}^2\rightarrow A T^2A\sim A_{23}T_{12}T_{34}A_{23}$, as discussed above. 
 
Equations for the bound state wave function $\Psi$, corresponding to  \eqs{modified-eq}, are derived by taking the residue of $X$ at the pole in the energy plane corresponding to the mass of the tetraquark. Defining the wave function components corresponding to the amplitudes of \eqs{amps1},
as
\be
\Psi=\sum_{aa'} \Psi_{aa'}  \eqn{Psi}
\ee
where
\begin{subequations}
\begin{align}
\Psi_{aa'} &= \Psi^T_{aa'} + \Psi^A_{aa'} \\[1mm]
\Psi^T_{aa'} &= \Psi^1_{aa'} + \Psi^2_{aa'},
\end{align}
\end{subequations}
the bound state equations corresponding to \eqs{modified-eq} are
\begin{subequations}
\label{Psi-eq}
\begin{align}
\Psi^A_{aa'}&= A_{aa'}(\Psi^2_{bb'}+\Psi^2_{cc'}),\\[1mm]
\Psi^1_{aa'}&= T^1_{aa'}(\Psi^T_{bb'}+\Psi^T_{cc'}),\\[1mm]
\Psi^2_{aa'}&=  T^2_{aa'}(\Psi_{bb'}+\Psi_{cc'}),
\end{align}
\end{subequations}
where $aa'\neq bb'\neq cc'  \neq aa'$.

As we mentioned above, \eq{Psi-eq} will be considered in full in a later publication; while here
we consider a simpler approximation, corresponding to setting $T^1_{aa'}=0$, in which case the kernels are given explicitly by \eqs{appr-Taa'}.
The equations for the $2q2\q$ amplitude $X$, are then obtained from \eqs{amps1}, \eq{X=Xaa'}, and  \eqs{modified-eq}, by setting ${\cal T}_{aa'}^1=0$, and therefore ${\cal T}_{aa'}^2={\cal T}_{aa'}$:
\begin{subequations}
\begin{align}
X &=\sum_{aa'} {\cal X}_{aa'}\\[1mm]
{\cal X}_{aa'}&={\cal T}_{aa'}+{\cal A}_{aa'}
\end{align}
\end{subequations}
where
\begin{subequations}
\label{simpler-eq}
\begin{align}
{\cal A}_{aa'}&= A_{aa'} + A_{aa'}\left({\cal T}_{bb'}+{\cal T}_{cc'}\right)
\\[1mm]
{\cal T}_{aa'}&= T_{aa'} + T_{aa'}\left({\cal X}_{bb'}+{\cal X}_{cc'}\right)
\end{align}
\end{subequations}
and $aa'\neq bb'\neq cc' \neq aa'$. 
Similarly, equations for the bound state wave function are
\begin{subequations}
\begin{align}
\Psi &=\sum_{aa'} \Psi_{aa'}\\[1mm]
\Psi_{aa'}&=\Psi^T_{aa'}+ \Psi^A_{aa'}
\end{align}
\end{subequations}
where 
\begin{subequations}
\label{Psi-simp}
\begin{align}
\Psi^A_{aa'}&= A_{aa'}(\Psi^T_{bb'}+\Psi^T_{cc'}),\\[1mm]
\Psi^T_{aa'}&=  T_{aa'}(\Psi_{bb'}+\Psi_{cc'}),
\end{align}
\end{subequations}
and $aa'\neq bb'\neq cc' \neq aa'$.


\begin{thebibliography}{10}%
\makeatletter
\providecommand \@ifxundefined [1]{%
 \@ifx{#1\undefined}
}%
\providecommand \@ifnum [1]{%
 \ifnum #1\expandafter \@firstoftwo
 \else \expandafter \@secondoftwo
 \fi
}%
\providecommand \@ifx [1]{%
 \ifx #1\expandafter \@firstoftwo
 \else \expandafter \@secondoftwo
 \fi
}%
\providecommand \natexlab [1]{#1}%
\providecommand \enquote  [1]{``#1''}%
\providecommand \bibnamefont  [1]{#1}%
\providecommand \bibfnamefont [1]{#1}%
\providecommand \citenamefont [1]{#1}%
\providecommand \href@noop [0]{\@secondoftwo}%
\providecommand \href [0]{\begingroup \@sanitize@url \@href}%
\providecommand \@href[1]{\@@startlink{#1}\@@href}%
\providecommand \@@href[1]{\endgroup#1\@@endlink}%
\providecommand \@sanitize@url [0]{\catcode `\\12\catcode `\$12\catcode
  `\&12\catcode `\#12\catcode `\^12\catcode `\_12\catcode `\%12\relax}%
\providecommand \@@startlink[1]{}%
\providecommand \@@endlink[0]{}%
\providecommand \url  [0]{\begingroup\@sanitize@url \@url }%
\providecommand \@url [1]{\endgroup\@href {#1}{\urlprefix }}%
\providecommand \urlprefix  [0]{URL }%
\providecommand \Eprint [0]{\href }%
\providecommand \doibase [0]{http://dx.doi.org/}%
\providecommand \selectlanguage [0]{\@gobble}%
\providecommand \bibinfo  [0]{\@secondoftwo}%
\providecommand \bibfield  [0]{\@secondoftwo}%
\providecommand \translation [1]{[#1]}%
\providecommand \BibitemOpen [0]{}%
\providecommand \bibitemStop [0]{}%
\providecommand \bibitemNoStop [0]{.\EOS\space}%
\providecommand \EOS [0]{\spacefactor3000\relax}%
\providecommand \BibitemShut  [1]{\csname bibitem#1\endcsname}%
\let\auto@bib@innerbib\@empty
\bibitem [{\citenamefont {Heupel}\ \emph {et~al.}(2012)\citenamefont {Heupel},
  \citenamefont {Eichmann},\ and\ \citenamefont {Fischer}}]{heupel}%
  \BibitemOpen
  \bibfield  {author} {\bibinfo {author} {\bibfnamefont {W.}~\bibnamefont
  {Heupel}}, \bibinfo {author} {\bibfnamefont {G.}~\bibnamefont {Eichmann}}, \
  and\ \bibinfo {author} {\bibfnamefont {C.~S.}\ \bibnamefont {Fischer}},\
  }\href {\doibase dx.doi.org/10.1016/j.physletb.2012.11.009} {\bibfield
  {journal} {\bibinfo  {journal} {Phys. Lett.}\ }\textbf {\bibinfo {volume}
  {B718}},\ \bibinfo {pages} {545} (\bibinfo {year} {2012})},\ \Eprint
  {http://arxiv.org/abs/1206.5129} {arXiv:1206.5129 [hep-ph]} \BibitemShut
  {NoStop}%
\bibitem [{\citenamefont {Popovici}\ and\ \citenamefont
  {Fischer}(2014)}]{popovici}%
  \BibitemOpen
  \bibfield  {author} {\bibinfo {author} {\bibfnamefont {C.}~\bibnamefont
  {Popovici}}\ and\ \bibinfo {author} {\bibfnamefont {C.~S.}\ \bibnamefont
  {Fischer}},\ }\href@noop {} {\  (\bibinfo {year} {2014})},\ \Eprint
  {http://arxiv.org/abs/1403.5900} {arXiv:1403.5900 [hep-ph]} \BibitemShut
  {NoStop}%
\bibitem [{\citenamefont {Khvedelidze}\ and\ \citenamefont
  {Kvinikhidze}(1992)}]{kvin4q}%
  \BibitemOpen
  \bibfield  {author} {\bibinfo {author} {\bibfnamefont {A.~M.}\ \bibnamefont
  {Khvedelidze}}\ and\ \bibinfo {author} {\bibfnamefont {A.~N.}\ \bibnamefont
  {Kvinikhidze}},\ }\href@noop {} {\bibfield  {journal} {\bibinfo  {journal}
  {Theor. Math. Phys.}\ }\textbf {\bibinfo {volume} {90}},\ \bibinfo {pages}
  {62} (\bibinfo {year} {1992})}\BibitemShut {NoStop}%
\bibitem [{\citenamefont {Kvinikhidze}\ and\ \citenamefont
  {Blankleider}(1994{\natexlab{a}})}]{piNN}%
  \BibitemOpen
  \bibfield  {author} {\bibinfo {author} {\bibfnamefont {A.~N.}\ \bibnamefont
  {Kvinikhidze}}\ and\ \bibinfo {author} {\bibfnamefont {B.}~\bibnamefont
  {Blankleider}},\ }\href {\doibase 10.1016/0375-9474(94)90959-8} {\bibfield
  {journal} {\bibinfo  {journal} {Nucl. Phys.}\ }\textbf {\bibinfo {volume}
  {A574}},\ \bibinfo {pages} {788} (\bibinfo {year} {1994}{\natexlab{a}})},\
  \Eprint {http://arxiv.org/abs/nucl-th/9402010} {arXiv:nucl-th/9402010}
  \BibitemShut {NoStop}%
\bibitem [{\citenamefont {Blankleider}\ and\ \citenamefont
  {Kvinikhidze}(2000)}]{overcount}%
  \BibitemOpen
  \bibfield  {author} {\bibinfo {author} {\bibfnamefont {B.}~\bibnamefont
  {Blankleider}}\ and\ \bibinfo {author} {\bibfnamefont {A.~N.}\ \bibnamefont
  {Kvinikhidze}},\ }\href@noop {} {\bibfield  {journal} {\bibinfo  {journal}
  {Phys. Rev. C}\ }\textbf {\bibinfo {volume} {62}},\ \bibinfo {pages} {039801}
  (\bibinfo {year} {2000})},\ \Eprint {http://arxiv.org/abs/nucl-th/9912003}
  {nucl-th/9912003} \BibitemShut {NoStop}%
\bibitem [{\citenamefont {Kvinikhidze}\ and\ \citenamefont
  {Blankleider}(1999{\natexlab{a}})}]{gaug1}%
  \BibitemOpen
  \bibfield  {author} {\bibinfo {author} {\bibfnamefont {A.~N.}\ \bibnamefont
  {Kvinikhidze}}\ and\ \bibinfo {author} {\bibfnamefont {B.}~\bibnamefont
  {Blankleider}},\ }\href {\doibase 10.1103/PhysRevC.60.044003} {\bibfield
  {journal} {\bibinfo  {journal} {Phys. Rev. C}\ }\textbf {\bibinfo {volume}
  {60}},\ \bibinfo {pages} {044003} (\bibinfo {year} {1999}{\natexlab{a}})},\
  \Eprint {http://arxiv.org/abs/nucl-th/9901001} {arXiv:nucl-th/9901001}
  \BibitemShut {NoStop}%
\bibitem [{\citenamefont {Kvinikhidze}\ and\ \citenamefont
  {Blankleider}(1999{\natexlab{b}})}]{gaug2}%
  \BibitemOpen
  \bibfield  {author} {\bibinfo {author} {\bibfnamefont {A.~N.}\ \bibnamefont
  {Kvinikhidze}}\ and\ \bibinfo {author} {\bibfnamefont {B.}~\bibnamefont
  {Blankleider}},\ }\href {\doibase 10.1103/PhysRevC.60.044004} {\bibfield
  {journal} {\bibinfo  {journal} {Phys. Rev. C}\ }\textbf {\bibinfo {volume}
  {60}},\ \bibinfo {pages} {044004} (\bibinfo {year} {1999}{\natexlab{b}})},\
  \Eprint {http://arxiv.org/abs/nucl-th/9901002} {arXiv:nucl-th/9901002}
  \BibitemShut {NoStop}%
\bibitem [{\citenamefont {Kvinikhidze}\ and\ \citenamefont
  {Blankleider}(1997)}]{gaug-spect}%
  \BibitemOpen
  \bibfield  {author} {\bibinfo {author} {\bibfnamefont {A.~N.}\ \bibnamefont
  {Kvinikhidze}}\ and\ \bibinfo {author} {\bibfnamefont {B.}~\bibnamefont
  {Blankleider}},\ }\href {\doibase 10.1103/PhysRevC.56.2973} {\bibfield
  {journal} {\bibinfo  {journal} {Phys. Rev. C}\ }\textbf {\bibinfo {volume}
  {56}},\ \bibinfo {pages} {2973} (\bibinfo {year} {1997})},\ \Eprint
  {http://arxiv.org/abs/nucl-th/9706052} {arXiv:nucl-th/9706052} \BibitemShut
  {NoStop}%
\bibitem [{\citenamefont {Fischer}\ \emph {et~al.}(2014)\citenamefont
  {Fischer}, \citenamefont {Kubrak},\ and\ \citenamefont
  {Williams}}]{Fischer:2014xha}%
  \BibitemOpen
  \bibfield  {author} {\bibinfo {author} {\bibfnamefont {C.~S.}\ \bibnamefont
  {Fischer}}, \bibinfo {author} {\bibfnamefont {S.}~\bibnamefont {Kubrak}}, \
  and\ \bibinfo {author} {\bibfnamefont {R.}~\bibnamefont {Williams}},\
  }\href@noop {} {\  (\bibinfo {year} {2014})},\ \Eprint
  {http://arxiv.org/abs/1406.4370} {arXiv:1406.4370 [hep-ph]} \BibitemShut
  {NoStop}%
\bibitem [{\citenamefont {Kvinikhidze}\ and\ \citenamefont
  {Blankleider}(1994{\natexlab{b}})}]{Kvinikhidze:1993bn}%
  \BibitemOpen
  \bibfield  {author} {\bibinfo {author} {\bibfnamefont {A.~N.}\ \bibnamefont
  {Kvinikhidze}}\ and\ \bibinfo {author} {\bibfnamefont {B.}~\bibnamefont
  {Blankleider}},\ }\href {\doibase 10.1016/0375-9474(94)90959-8} {\bibfield
  {journal} {\bibinfo  {journal} {Nucl. Phys.}\ }\textbf {\bibinfo {volume}
  {A574}},\ \bibinfo {pages} {788} (\bibinfo {year} {1994}{\natexlab{b}})},\
  \Eprint {http://arxiv.org/abs/nucl-th/9402010} {arXiv:nucl-th/9402010}
  \BibitemShut {NoStop}%
\end{thebibliography}
%

\end{document}